\documentclass{aa}

\usepackage{graphicx}

\usepackage{txfonts}
\usepackage{gensymb}
\usepackage{hyperref}

\usepackage{natbib}
\bibpunct{(}{)}{;}{a}{}{,}

\usepackage[normalem]{ulem}

\begin{document}

   \title{{Periodicities in radio emissions from Jupiter{'s} magnetosphere and consequences for radio emissions from star--exoplanet systems}}
   \subtitle{}
   \titlerunning{{Periodicities in radio emissions from the Jupiter's magnetosphere}}

   \author{C. K. Louis\inst{1,2}
          \and
          A. Loh\inst{1,2}
          \and
          P. Zarka\inst{1,2}
          \and
          L. Lamy\inst{1,2,3}
          \and
          E. Mauduit\inst{1,2}
          \and 
          J. Girard\inst{1,2}
          \and \\
          J.--M. Grie\ss meier\inst{2,4}
          \and
          B. Cecconi\inst{1,2}
          \and
          Q. Nénon\inst{5}
          \and
          S. Corbel\inst{6}
          }

   \institute{LIRA, Observatoire de Paris, Université PSL, Sorbonne Université, Université Paris Cité, CY Cergy Paris Université, CNRS, 92190 Meudon, France\\
              \email{corentin.louis@obspm.fr}
              \and ORN, Observatoire Radioastronomique de Nan\c{c}ay, Observatoire de Paris, CNRS, Univ. PSL, Univ. Orl\'{e}ans, F--18330 Nan\c{c}ay, France
              \and Aix Marseille Université, CNRS, CNES, LAM, Marseille, France
              \and LPC2E -- Universit\'e d'Orl\'eans/CNRS, France
              \and LATMOS, CNRS -- Sorbonne Université -- CNES, Paris, France
              \and Université Paris Cité and Université Paris Saclay, CEA, CNRS, AIM, 91190 Gif--sur--Yvette, France 
             } 

   \date{}

  \abstract 
  {The search for radio signals from exoplanets or star--planet interactions is a {topic of major scientific interest}, as it is likely the best way {to detect and measure} a planetary magnetic field {and, therefore, to probe} the inner structure of exoplanets. {However, detecting these radio emissions is challenging, since they are anisotropic by nature, sporadic, and of low intensity because of their great distances, and because the sky cannot be monitored continuously.}}
   {The aim of this article is to demonstrate the relevance of using statistical tools to detect periodic radio signals in unevenly spaced observations and to identify the implications of the measured period.}
   {{The} identification of periodic radio signal{s} was achieved here through a Lomb--Scargle analysis. The technique was first applied {to} simulated astrophysical observations with controlled simulated noise. This allowed {us} to characterise the origin of spurious detection peaks in the resulting periodograms --as well as {to identify} peaks {corresponding} to real periods in the studied system-- {and to} {combination} or {beat} periods.}
  {The method was validated using a real signal, with $\sim 1400$~hours of data from observations of Jupiter's radio emissions by the NenuFAR radio telescope over more than six years, in order to detect {the periodicities of} Jovian radio emissions (auroral and induced by the Galilean moons).}
   {We demonstrate with the simulation that the Lomb--Scargle periodogram allows {us} to correctly identify periodic radio signal{s}, even in a diluted signal. On real measurements, it correctly detects the rotation period of the strong signal produced by Jupiter and the {beat} period of the emission triggered by the interaction between Jupiter and its Galilean moon Io, but also possibly weaker signal{s,} such as {those} produced by the interaction between Jupiter and Europa or between Jupiter and Ganymede. It is important to note that secondary peaks in the Lomb--Scargle periodogram {appear} at the {beat} and {combination} periods {among} all the detected signal periodicities (i.e. real signals, {but also periodicities due to regular observation intervals}). These secondary peaks can then be used to strengthen the detection of weak signals. Finally, the importance of {the number of observation windows used in the Lomb--Scargle analysis is discussed,} {as well as the data's} time and frequency {resolutions in} increas{ing} its efficiency.}


   \maketitle

\section{Introduction}

The most successful way to image {stellar} magnetic fields is to use spectro--imaging techniques such as Zeeman Doppler imaging (ZDI), which can detect a stellar magnetic field down to 0.8 G \citep{2022MNRAS.514.4300B}. However, in the case of exoplanetary magnetic fields, the spectrum of the exoplanet is usually combined with that of the star, making it difficult to extract the exoplanet's polarimetric signature, and the photon noise can severely limit the possibility of detecting this potentially weak polarimetric signal. Finally, the exoplanet magnetic field {might} also not be strong enough to produce a sufficient Zeeman effect on the atom{s} or molecule{s} of the exoplanet's atmosphere, the {composition} and temperature of which differ from those of a star \citep[for a low limit for low temperatures, see, e.g.][ which presents the detection of a magnetic field of $5$~kG for the brown dwarf LSR J1835+3259 of spectral type M8.5V]{2017ApJ...847...60K}. Consequently, ZDI {is} unable to characterise exoplanetary magnetic fields.

Indirect detections of exoplanetary magnetic fields have been achieved using techniques based on planet-modulated chromospheric emission \citep{2019NatAs...3.1128C} or neutral atomic hydrogen absorption during transit \citep{2022NatAs...6..141B}.
{Another} way to study and characterise exoplanetary magnetic fields is through the observation of exoplanetary auroral radio emissions. These emissions are produced by the cyclotron maser instability \citep[CMI, ][]{2006A&ARv..13..229T} at all magnetised planet{s} in our Solar System \citep{1998JGR...10320159Z} and are {particularly} well known and studied in situ on Earth, Jupiter, and Saturn \citep{1979ApJ...230..621W, 1984JGR....89.2831L,1984PhFl...27..247L,  1985SSRv...41..215W, 1986JGR....9113569P, 2010GeoRL..3712104L, 2010GeoRL..3719105M, 2011pre7.conf...75K, 2017GeoRL..44.4439L, 2018GeoRL..45.9408L, 2023JGRA..12831985L, 2023_Collet_PRE9, 2024JGRA..12932422C}. They occur at or near the fundamental frequency of the local electron cyclotron frequency:
\begin{equation}
    f_\mathrm{ce} = \frac{eB}{2 \pi m_\mathrm{e}}
    \label{eq:fce}
,\end{equation}
with $e$ and $m_\mathrm{e}$ being the charge and mass of an electron and $B$ the local magnetic-field amplitude. As the magnetic-field amplitude decreases with increasing altitude above the atmosphere, auroral radio emissions span a broad range of frequencies along the magnetic-field lines, provided the ratio between the plasma frequency  and the cyclotron frequency is less than $\sim0.1$ {\citep[from theoretical work on the CMI and from in situ measurements by Juno at Jupiter; ][]{2001P&SS...49.1137Z, 2002PhPl....9.2816C, 2013JGRA..118.7036L,2023JGRA..12831985L,2024JGRA..12932422C,2025_Collet_GRL}}. The minimal frequency for the emission is therefore reached at high altitudes (several body radii), while the maximal frequency corresponds to the maximal cyclotron frequency near the planetary surface.

Consequently, detecting these radio emissions provides information about the local magnetic field. For example, in our Solar System, Jovian auroral radio emissions were first detected (above the ionospheric cut-off at $10$~MHz) in 1955 by \citet{1955JGR....60..213B}, making Jupiter's magnetosphere the first to be identified using the radio radiation. Jovian auroral radio emissions extend from a few kilohertz (kilometric range) up to 40 megahertz (decametric range), corresponding to a magnetic field with a maximal amplitude of about $15$~G \citep{2022JGRE..12707055C}.

These radio emissions are highly anisotropic. They are produced along the edges of a hollow cone, $\sim 1--2 \degree$ thick, with an opening angle relative to the local magnetic-field vector, $\vec{B,}$ varying from $\sim 70 \degree$ to $90 \degree$ depending on the type of electron distribution function and the emission frequency \citep{1986PhFl...29.2919P, 2006A&ARv..13..229T, 2008GeoRL..3513107H}. Consequently, the visibility of these emissions strongly depends on the observer's position in the reference frame of the emitting body \citep{2023pre9.conf03091L}. For instance, in the case of emissions induced by the moon Io \citep{1964Natur.203.1008B}, a terrestrial observer located approximately in the plane of the Jovian equator can only see the emissions when Io is in quadrature \citep{2017A&A...604A..17M}. Saturn-like auroral radio emissions (fast rotator, sensitive to stellar wind) are mostly visible for an observer in the morning sector, and hence also in quadrature \citep{2017pre8.conf..171L, 2023pre9.conf03091L, 2008JGRA..113.7201L, 2013JGRA..118.7019K, 2019P&SS..17804711N}. For Earth-like auroral radio emissions (highly sensitive to stellar wind) the maximal chance of detection is for an observer located in the nightside sector \citep{2022JGRA..12730449W, 2023pre9.conf03091L, 2023pre9.conf03088L}. Finally, Jupiter-like auroral {(non-satellite-induced)} radio emissions (strongly magnetised and fast rotator planet, partially sensitive to stellar winds) do not require any preferred position of observation for most of {their} components \citep{2021JGRA..12629780Z, 2021JGRA..12629435L, 2023pre9.conf03091L, 2023pre9.conf03094B}.

Another important characteristic of auroral radio emissions is their strong polarisation, which is almost 100\%. These emissions are produced {in} the extraordinary mode (so-called R-X mode) and are therefore circularly or elliptically polarised in the right-handed (RH) sense relative to the magnetic field  at the source. The observed polarisation thus depends on the magnetic hemisphere: it is RH when the ($\vec{B}$,$\vec{k}$) angle is acute (with $\vec{k}$ being the radio-wave vector) and left-handed (LH) when the ($\vec{B}$,$\vec{k}$) angle is obtuse.
However, these {signals} are weak at stellar distances and highly sporadic. To detect these weak astrophysical signals \citep{2018A&A...618A..84Z}, it is crucial to use prediction tools (e.g. ExPRES, \citet{2019A&A...627A..30L}; PALANTIR, \citet{2023pre9.conf03092M}; phase prediction\footnote{\url{https://github.com/zhangxiang-planet/orbital_coverage}}, \citet{2025_ZHANG_Xiang_HD189733}). Another problem is that the detections published so far have been individual bursts \citep{2021A&A...648A..13C, 2021NatAs...5.1233C,2023A&A...670A.124C, 2024NatAs...8.1359C, 2025NatAs...Tasse, 2021A&A...645A..59T,2023pre9.conf04048T, 2024A&A...688A..66T, 2020NatAs...4..577V}; thus, it is not possible to draw conclusions about their origin in the observed stellar systems (either from the star itself, an exoplanet, or a star--planet interaction).

With sufficient observation hours over an extended period using sensitive radio telescopes \citep{2019A&A...624A..40T}, it should in principle become possible to search for periodicity in the signal, instead of searching for individual weak and sporadic radio emissions. However, a significant challenge arises from the uneven spacing of these observations over time. Due to the limited observability of the sources (not 24 hours a day), observation biases (at $24$~hours because of the radio frequency interferences that {pollute} the observations, and at $23.93$~hours, which is the {sidereal} period), and high observing pressure on giant telescopes such as NenuFAR \citep{Zarka_2020_NenuFAR}, regular observations are impractical. Consequently, periodicity search techniques capable of handling unevenly spaced observations are required, and the Lomb--Scargle (LS) periodogram such as the one provided by the Astropy Python package \citep{astropy:2013, astropy:2018, astropy:2022, 2012cidu.conf...47V, 2015ApJ...812...18V} is an excellent option for this purpose. The LS periodogram has, for instance, proven to be successful {in tracking} the double radio period of Saturn's kilometric radiation \citep[e.g.][]{2017pre8.conf..171L}, the period of Jovian quasi-periodic bursts \citep{2011JGRA..116.3204K}, {and the brown-dwarf period \citep[e.g.][]{2018ApJS..237...25K, 2023A&A...675L...6V, 2024A&A...682A.170B}.}

In this work, a practical test was conducted on real NenuFAR data. To understand how the LS periodogram performs on NenuFAR data, a first test was done on a simulated signal (see Section~\ref{sec:simu}). Initially, a sine wave with a known periodicity and random observation gaps is used. Next, a similar simulation was produced, but this time the observation windows and gaps between observations were controlled to study the impact of observation regularity on the LS periodogram, hence mimicking real observing conditions. Finally, the signal was embedded in random noise with a normal distribution, and we studied how varying the signal-to-noise ratio affects the detection of the underlying periodic signal.
In Section~\ref{sec:real_signal}, the LS periodogram is applied to real observations of Jupiter's radio emissions acquired over a six-year interval with the NenuFAR radio telescope to analyse the periodicities {detected}.
Finally, this study is summarised and discussed in Section~\ref{sec:summary} to highlight the constraints and limitations of using this technique to detect radio emissions by searching for signal periodicity.

\section{Simulations}
\label{sec:simu}

   \begin{figure*}
   \resizebox{\hsize}{!}
            {\includegraphics[width=0.6\textwidth]{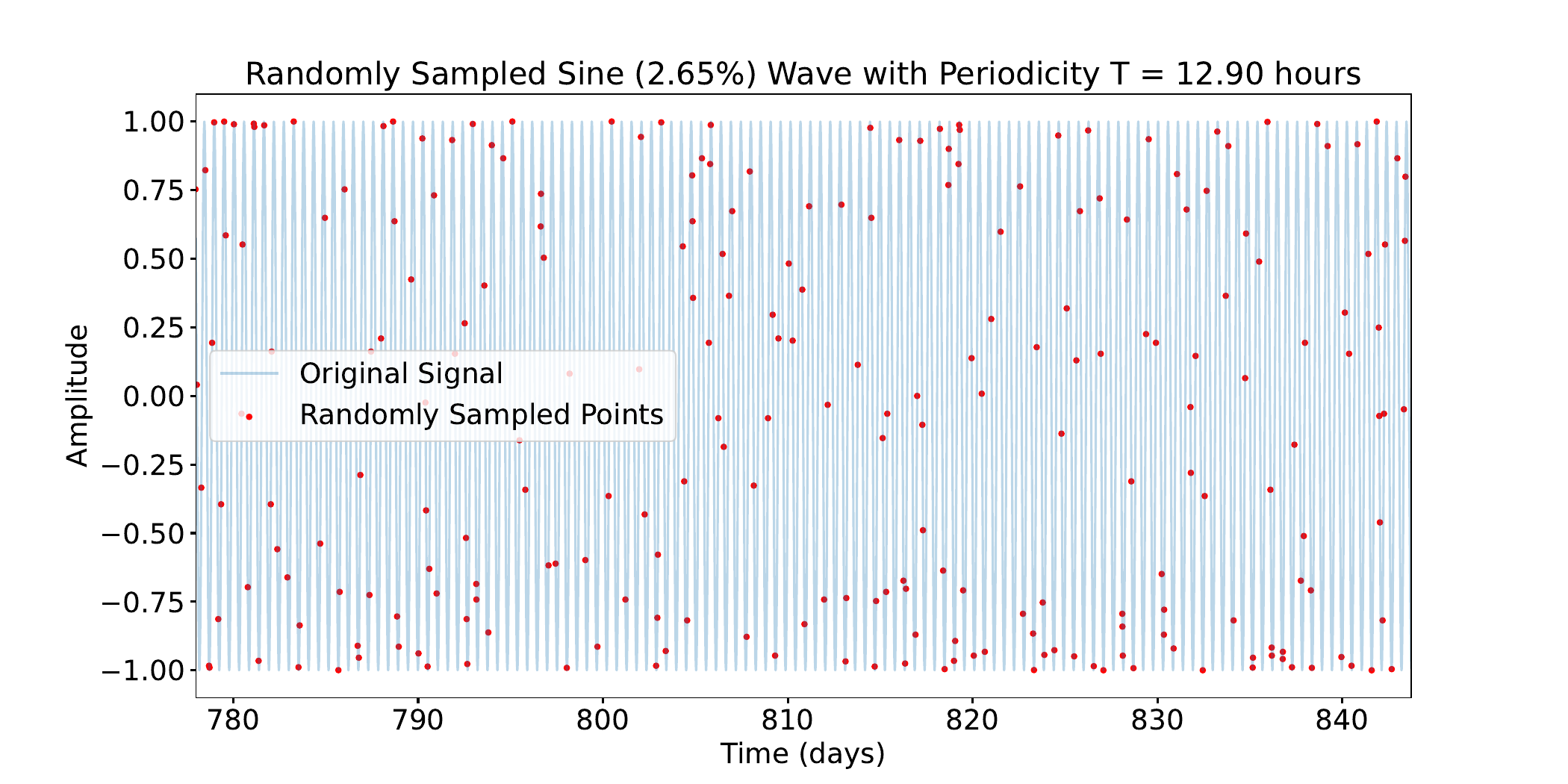}
            \includegraphics[width=0.4\textwidth]{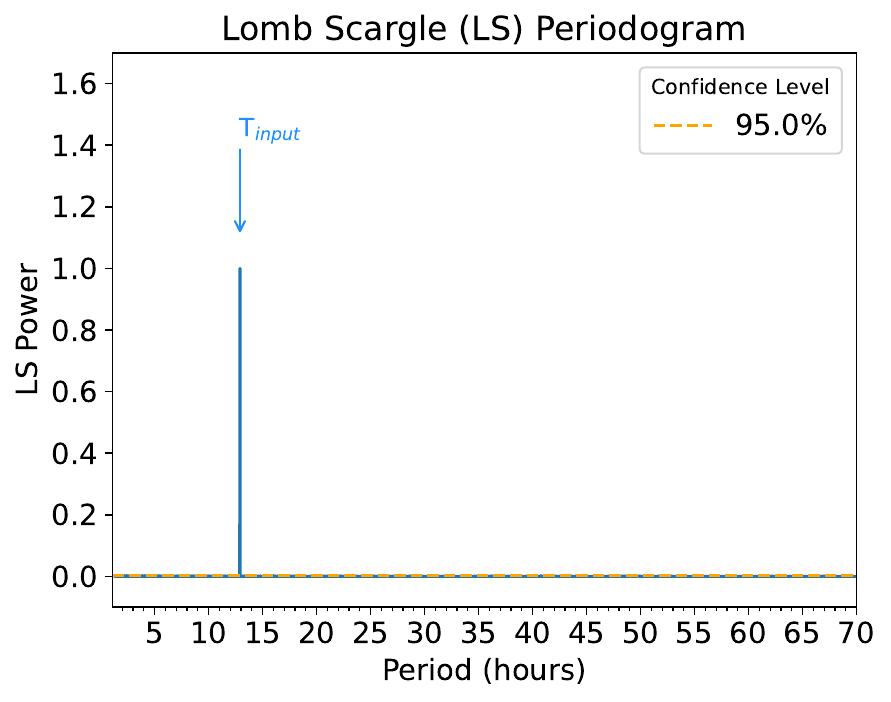}}
      \caption{(left) Randomly spaced sinusoidal wave with period T$_\mathrm{sine} = 12.90$~h. Only a few days of the entire five-year interval are shown in order to show the sinusoidal signal clearly. (right) Corresponding LS periodogram. The $95$\% confidence level (based on a randomisation test) is indicated for the LS periodogram. It is very low in this case as only signal is given (no noise).}
         \label{fig:random_spaced_sinus}
   \end{figure*}

   In the first simulation (see Figure \ref{fig:random_spaced_sinus}, left), a sinusoidal signal wave was created (in blue) with a periodicity of T$_\mathrm{sine} = 12.90~$~hours (with an amplitude between $-1$ and $1$; in preparation for the analysis given in Section \ref{sec:real_signal} of the data in circular polarisation, which therefore has a ratio of Stokes parameters V over I between $-1$ and $1$) over five years (only a few tens of days are shown for the sake of readability) with a time resolution of $600$~seconds. {A random selection of $2.65$\% of the data points in the time series is then made for analysis}. By {anticipating} the real signal, a periodicity of $12.9$~hours was chosen, as it is close to the real physical periodicity investigated in Section \ref{sec:real_signal}. The {percentage} of kept signal was chosen on the basis of the percentage of data actually observed, which we analyse in Section \ref{sec:real_signal}. The corresponding LS periodogram is shown on the right of Figure \ref{fig:random_spaced_sinus}. The highest peak (by far) in the periodogram is located at $12.9$~hours. The $95$\% confidence level, based on a randomisation test, is also shown. Our randomisation test consists of taking the {time series} under consideration, shuffling the values randomly over time and calculating the resulting LS periodogram. For this test to have sufficient statistical validity, it was repeated 1000 times. The confidence levels therefore indicate the percentage of times the highest peak reaches this power value. {This is equivalent to the false-alarm probability (FAP) using the bootstrap method in the LS Astropy Python package.} In the case of this first simulation, the $95$\% confidence level is obviously very low, as there is only signal given as input; moreover, the portion of kept signal is completely random.

   \begin{figure*}
   \resizebox{\hsize}{!}
            {\includegraphics[width=0.6\textwidth]{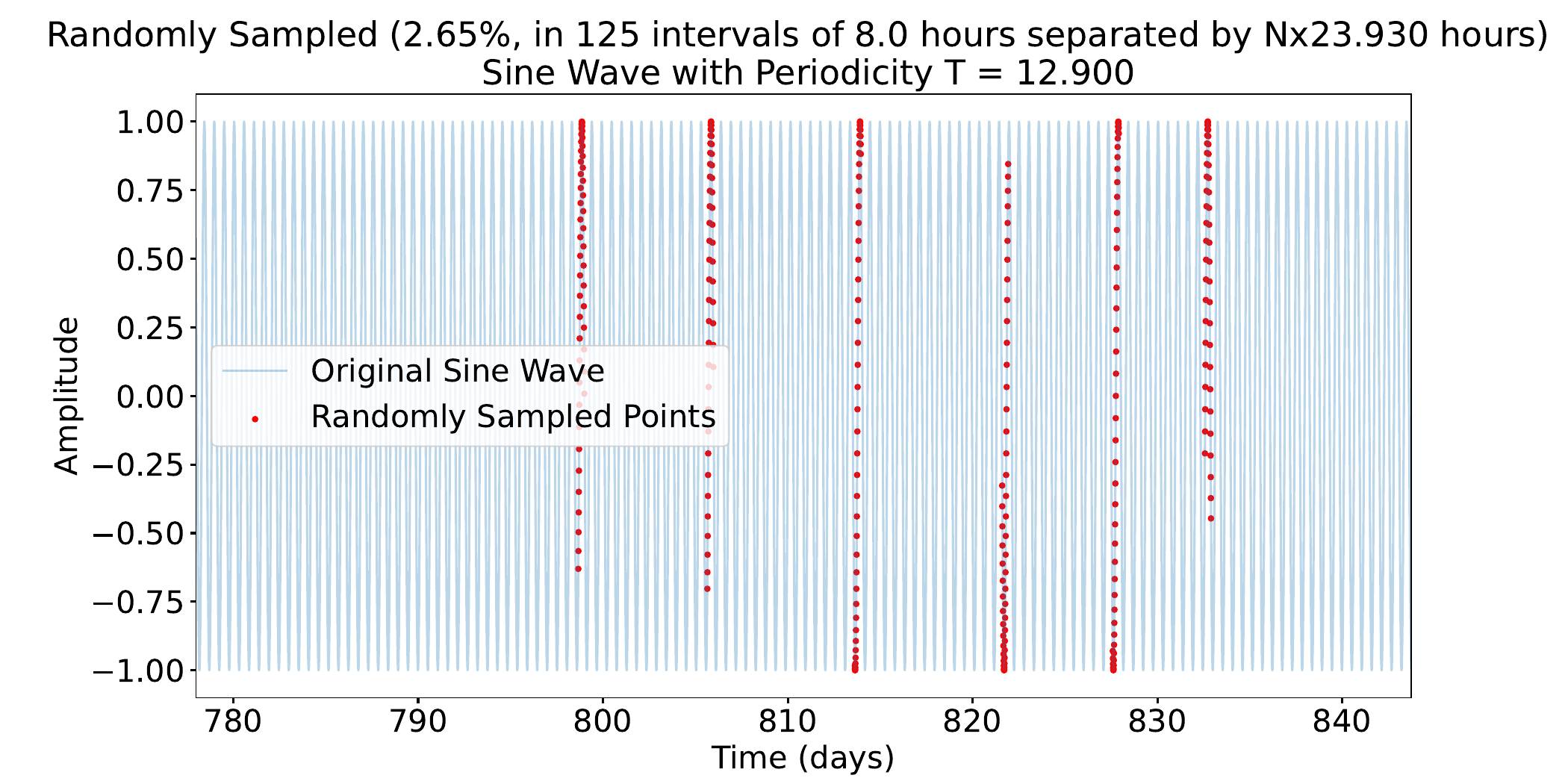}
            \includegraphics[width=0.4\textwidth]{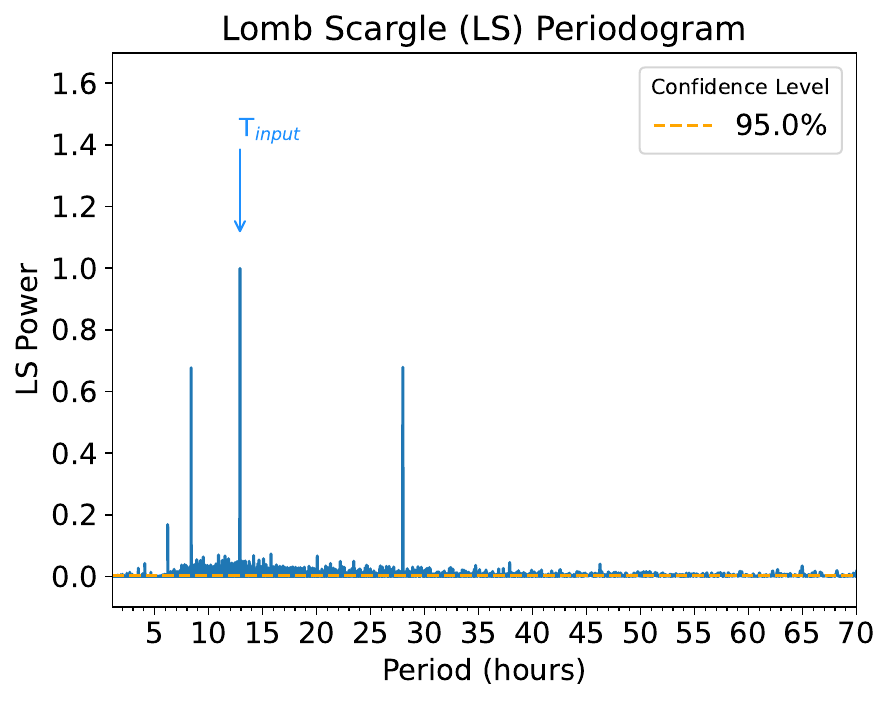}}
      \caption{(left) Semi-regularly spaced sinusoidal wave with period T$_\mathrm{sine} = 12.90$~h. The samples are gathered into $125$~intervals of eight hours, spaced by $\mathrm{N} \times 23.93$~hours (with $\mathrm{N} = 1, 2, 3, 4, ...$). (right) Corresponding LS periodogram. {The} $95$\% confidence level is also indicated. It is also very low in this case as only {a} signal is given (no noise).}
         \label{fig:regularly_spaced_sinus}
   \end{figure*}

   \begin{figure*}
      \resizebox{\hsize}{!}
               {\includegraphics[width=0.6\textwidth]{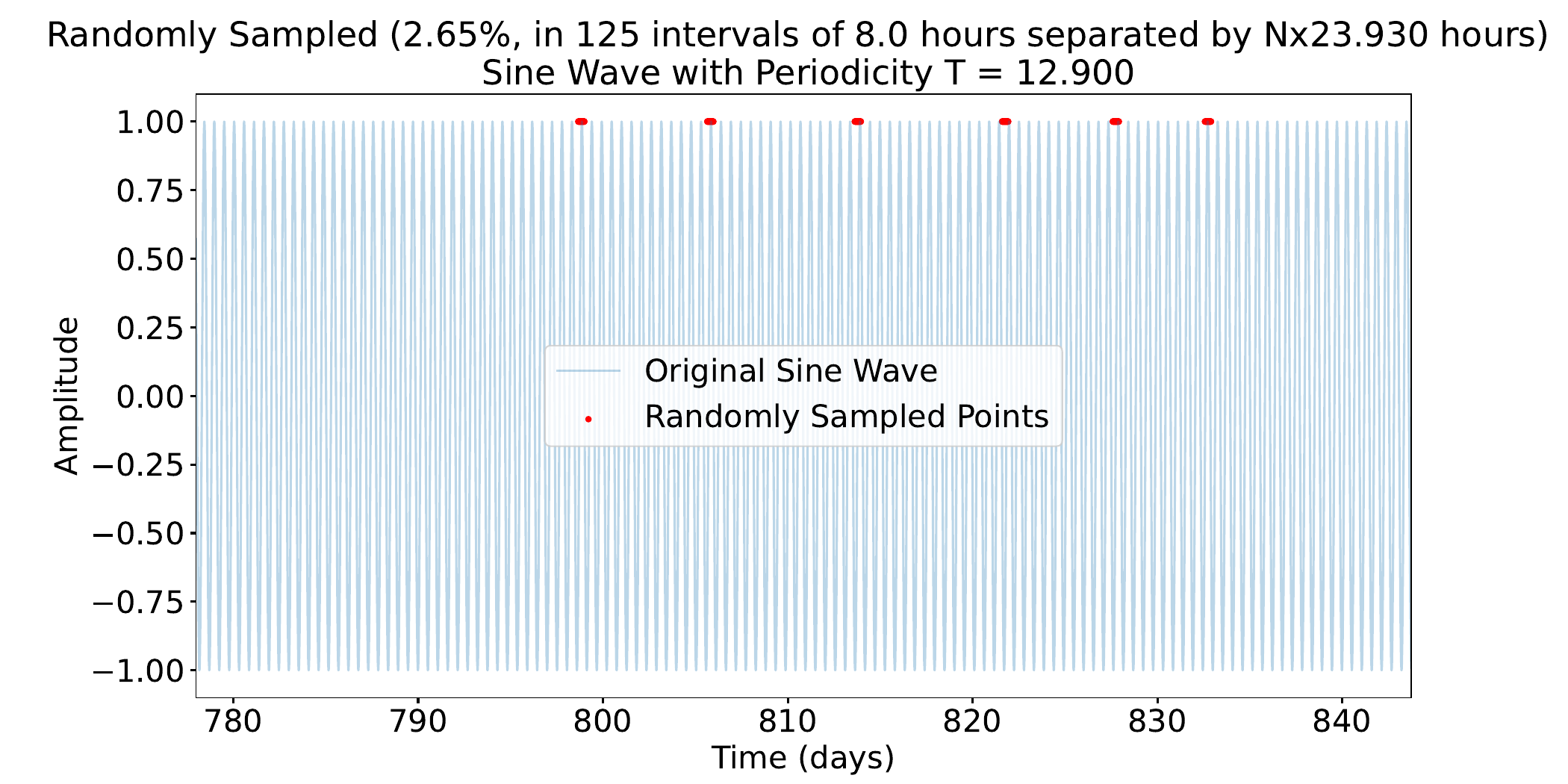}
               \includegraphics[width=0.4\textwidth]{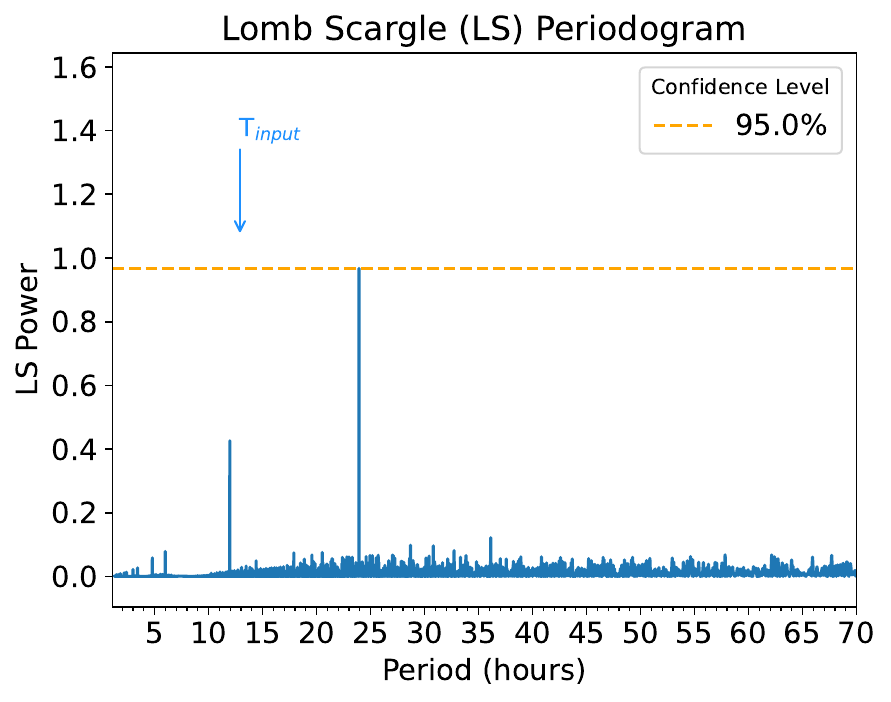}}
         \caption{Same as Figure~\ref{fig:regularly_spaced_sinus}, but with all values of the sinusoid at 1 in order to show the effect of the windowing on
         the LS periodogram. {The} $95$\% confidence level is also indicated. As all values are saturated to 1, randomly shuffling the values does not change the LS analysis, and the $95$\% confidence level is at the values of the highest peaks (due to the {sidereal} {periodicity} at $\mathrm{N} \times 23.93$ hours).}
       \label{fig:regularly_spaced_sinus_saturated_1}
      \end{figure*}

In Figure \ref{fig:regularly_spaced_sinus}, we show the simulation of the same sinusoidal wave (i.e. with a T$_\mathrm{sine} = 12.9$~h periodicity and a time resolution of $600$~seconds), but this time more realistic values were used to select the kept signals.{} $2.65$\% of the kept {signal was} gathered {into} $125$~{intervals} of eight hours separated by  $\mathrm{N} \times 23.93$~hours (with $\mathrm{N} = 1, 2, 3, 4,...$), i.e. the sidereal day. These values were chosen on the basis of the real {observational} values of the target studied in Section \ref{sec:real_signal}. The corresponding LS periodogram is shown on the right of Figure \ref{fig:regularly_spaced_sinus}. The highest peak is still located at $12.9$~hours, but this time several other peaks are also visible, mainly at $27.99$~hours, $8.38$~hours, and $6.21$~hours. 

In Figure \ref{fig:regularly_spaced_sinus_saturated_1}, the effect of the windowing on the LS periodogram is studied. In this figure, the same observation intervals {as} in Figure \ref{fig:regularly_spaced_sinus} are taken, but all values are equal to one (left panel). In the LS periodogram (right panel), the input sinusoidal periodicity is not visible anymore. However, two {major} peaks are detected at $23.93$~hours and $11.97$~hours.

\begin{figure*}[]
   \resizebox{\hsize}{!}
            {\includegraphics[width=0.6\textwidth]{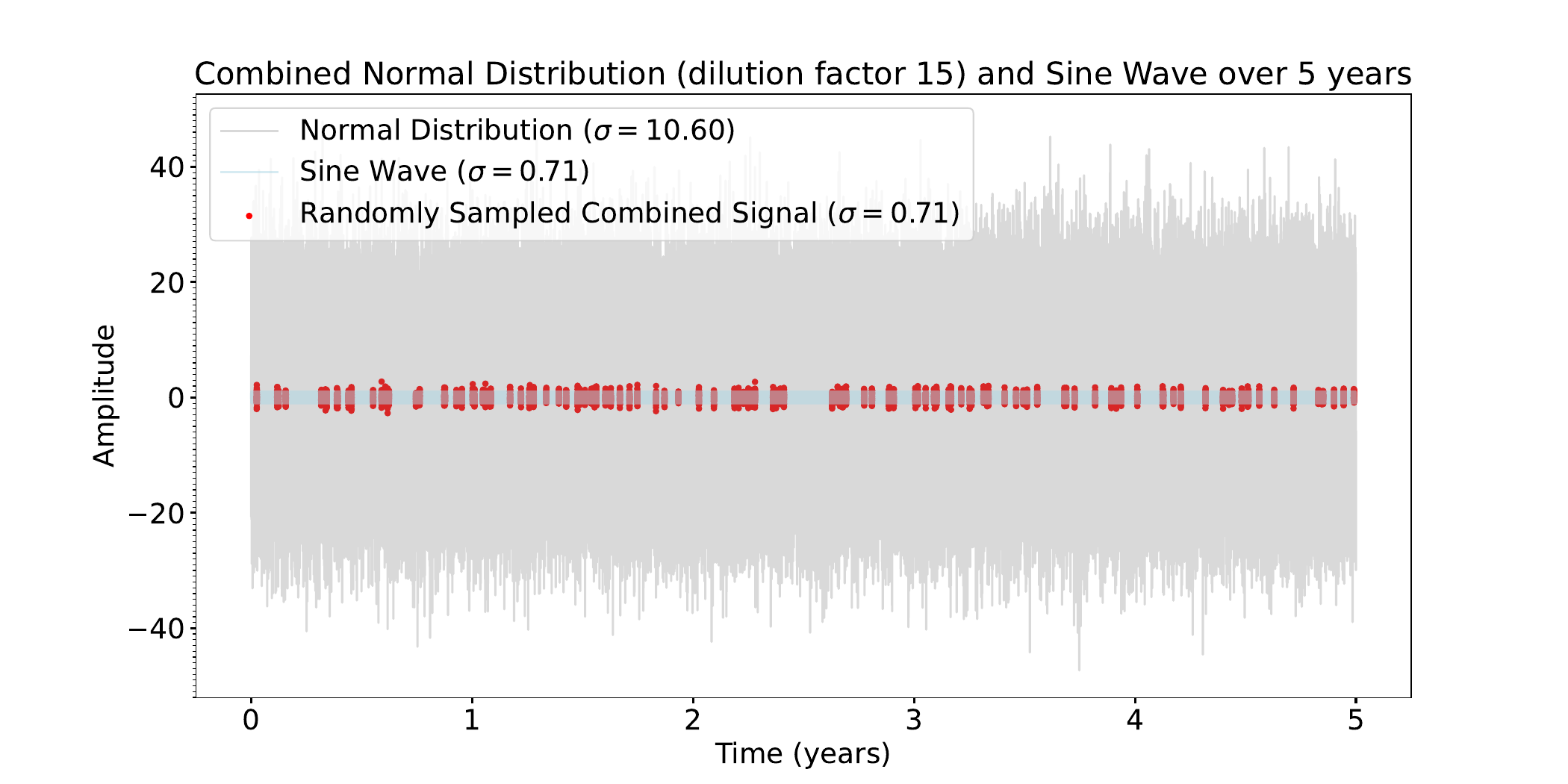}
            \includegraphics[width=0.4\textwidth]{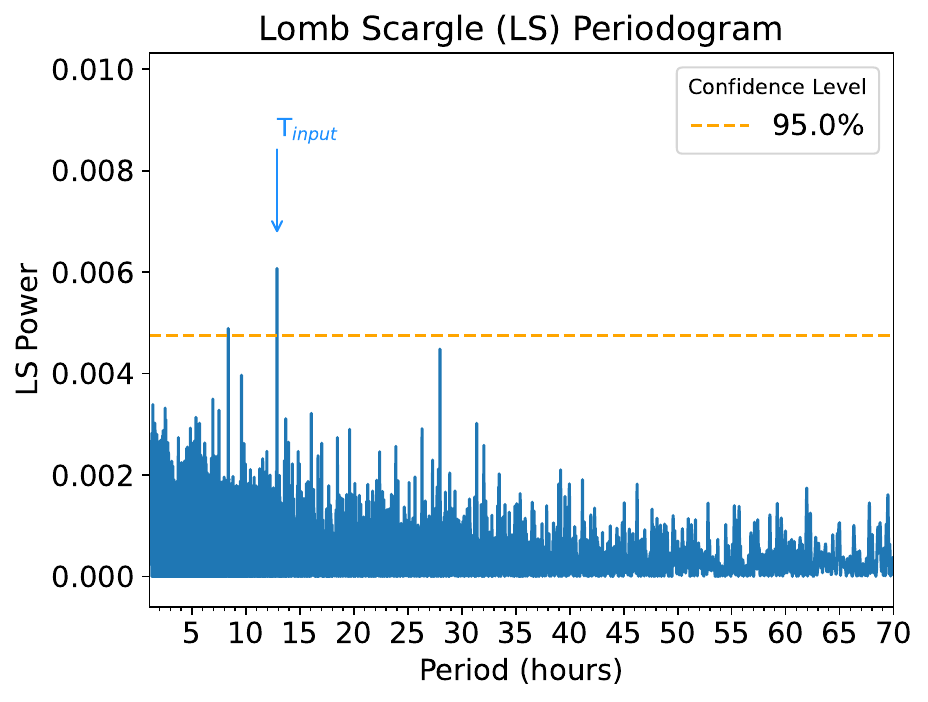}}
      \caption{(left panel) Semi-regularly spaced sinusoidal wave with period T$_\mathrm{sine} = 12.90$~h ({light blue}), normal distribution with a standard deviation of $\sigma = K \times \sigma_{sin}$ with $K=15$ ({light grey}), and combined signal with higher signal-to-noise ratio (red points). The samples are gathered into $125$~intervals of eight hours spaced by $\mathrm{N} \times 23.93$~hours. (right panel) Corresponding LS periodogram. The
      $95$\% confidence level is also shown in the LS periodogram.
      }
         \label{fig:regularly_spaced_sinus_diluted}
   \end{figure*}

To identify the origin of these different peaks, the possible {combinations between different periodicities} need to be studied. The two major combinations are (i) the difference between the two periodicities (in frequency), i.e. the {beat} period:
   \begin{equation}
      f_\mathrm{{beat}} = |f_1 - f_2|
   ,\end{equation}
which corresponds in time periodicity to
   \begin{equation}
      T_\mathrm{{beat}} = \frac{1}{f_\mathrm{{beat}}} =  \frac{T_1 \times T_2}{|T_1-T_2|}
   ,\end{equation}
and (ii) the sum of the two periodicities (in frequency), i.e. the combination period: 
\begin{equation}
   f_\mathrm{combination} = f_1 + m \times f_2
,\end{equation}
which corresponds in time periodicity to
   \begin{equation}
      T_\mathrm{combination} = \frac{1}{f_\mathrm{combination}} = \frac{1}{ 1/T_1 + m \times 1/T_2}
   ,\end{equation}
{with $m=1,2,...$.}

In our case, the simulation introduces three possible periodicities: $T_{sine} = 12.9$~h; the interval between two observations, $T_{gap} =  23.93$~h; {and} the length of the observations, $T_{obs} = 8$~h. {$T_\mathrm{obs}$ cannot really be considered as a periodicity since it is the duration of each observation, though it does have an effect on the windowing.} Several combinations are thus possible{, which are {summarised} in Table~\ref{tab:possible_periodicities_simulations}. Therefore, the peaks observed in Figure \ref{fig:regularly_spaced_sinus}
can be explained by (i) the {beat} period between T$_\mathrm{sine}   $ and T$_\mathrm{gap}$, and (ii) the combination period between $T_\mathrm{sine}$
and $T_\mathrm{gap}$, for $\mathrm{m} = 1$  and $\mathrm{m} = 2$.

\begin{table}[]
    \centering
    \caption{{Possible fundamental, beat, and combination periods that could be expected due to the sine, observed, and gap periods in our simulated signal.}}
    \label{tab:possible_periodicities_simulations}
    \small
\begin{tabular}{c|ccc}
\cline{2-4}
                                                                                   & \multicolumn{1}{c|}{T$_\mathrm{sine}$}                                     & \multicolumn{1}{c|}{T$_\mathrm{Obs}$}                                       & \multicolumn{1}{c|}{T$_\mathrm{gap}$}                                       \\ \hline
\multicolumn{1}{|c|}{\begin{tabular}[c]{@{}c@{}}Period\\ Half Period\end{tabular}} & \multicolumn{1}{c|}{\begin{tabular}[c]{@{}c@{}}12.9\\ 6.45\end{tabular}}   & \multicolumn{1}{c|}{\begin{tabular}[c]{@{}c@{}}8\\ 4\end{tabular}}          & \multicolumn{1}{c|}{\begin{tabular}[c]{@{}c@{}}23.93\\ 11.97\end{tabular}}  \\ \hline
\multicolumn{1}{|c|}{\begin{tabular}[c]{@{}c@{}}Beat\\ periods\end{tabular}}       &                                                                            &                                                                             &                                                                             \\ \hline
\multicolumn{1}{|c|}{T$_\mathrm{Sine}$}                                            & \multicolumn{1}{c|}{\begin{tabular}[c]{@{}c@{}}N/A\\ 12.90\end{tabular}}  & \multicolumn{1}{c|}{\begin{tabular}[c]{@{}c@{}}21.06\\ 5.798\end{tabular}}  & \multicolumn{1}{c|}{\begin{tabular}[c]{@{}c@{}}27.99\\ 165.1\end{tabular}}  \\ \hline
\multicolumn{1}{|c|}{T$_\mathrm{Obs}$}                                             & \multicolumn{1}{c|}{\begin{tabular}[c]{@{}c@{}}21.06\\ 33.29\end{tabular}} & \multicolumn{1}{c|}{\begin{tabular}[c]{@{}c@{}}N/A\\ 8.0\end{tabular}}     & \multicolumn{1}{c|}{\begin{tabular}[c]{@{}c@{}}12.02\\ 24.14\end{tabular}}  \\ \hline
\multicolumn{1}{|c|}{T$_\mathrm{gap}$}                                             & \multicolumn{1}{c|}{\begin{tabular}[c]{@{}c@{}}27.99\\ 8.830\end{tabular}} & \multicolumn{1}{c|}{\begin{tabular}[c]{@{}c@{}}12.018\\ 4.803\end{tabular}} & \multicolumn{1}{c|}{\begin{tabular}[c]{@{}c@{}}N/A\\ 23.93\end{tabular}}   \\ \hline
\multicolumn{1}{|c|}{\begin{tabular}[c]{@{}c@{}}Combination\\ periods\end{tabular}}   &                                                                            &                                                                             &                                                                             \\ \cline{1-1}
\multicolumn{1}{|c|}{n=1}                                                          & \multicolumn{1}{l}{}                                                       & \multicolumn{1}{l}{}                                                        & \multicolumn{1}{l}{}                                                        \\ \hline
\multicolumn{1}{|c|}{T$_\mathrm{Sine}$}                                            & \multicolumn{1}{c|}{\begin{tabular}[c]{@{}c@{}}6.45\\ 8.6\end{tabular}}    & \multicolumn{1}{c|}{\begin{tabular}[c]{@{}c@{}}4.938\\ 7.142\end{tabular}}  & \multicolumn{1}{c|}{\begin{tabular}[c]{@{}c@{}}8.382\\ 10.16\end{tabular}}  \\ \hline
\multicolumn{1}{|c|}{T$_\mathrm{Obs}$}                                             & \multicolumn{1}{c|}{\begin{tabular}[c]{@{}c@{}}4.938\\ 6.107\end{tabular}} & \multicolumn{1}{c|}{\begin{tabular}[c]{@{}c@{}}4.0\\ 5.333\end{tabular}}    & \multicolumn{1}{c|}{\begin{tabular}[c]{@{}c@{}}5.996\\ 6.854\end{tabular}}  \\ \hline
\multicolumn{1}{|c|}{T$_\mathrm{gap}$}                                             & \multicolumn{1}{c|}{\begin{tabular}[c]{@{}c@{}}8.382\\ 12.42\end{tabular}} & \multicolumn{1}{c|}{\begin{tabular}[c]{@{}c@{}}5.996\\ 9.589\end{tabular}}  & \multicolumn{1}{c|}{\begin{tabular}[c]{@{}c@{}}11.97\\ 15.95\end{tabular}}  \\ \hline
\multicolumn{1}{|c|}{n=2}                                                          &                                                                            &                                                                             &                                                                             \\ \hline
\multicolumn{1}{|c|}{T$_\mathrm{Sine}$}                                            & \multicolumn{1}{c|}{\begin{tabular}[c]{@{}c@{}}4.3\\ 6.45\end{tabular}}    & \multicolumn{1}{c|}{\begin{tabular}[c]{@{}c@{}}3.053\\ 4.938\end{tabular}}  & \multicolumn{1}{c|}{\begin{tabular}[c]{@{}c@{}}6.208\\ 8.382\end{tabular}}  \\ \hline
\multicolumn{1}{|c|}{T$_\mathrm{Obs}$}                                             & \multicolumn{1}{c|}{\begin{tabular}[c]{@{}c@{}}3.571\\ 4.938\end{tabular}} & \multicolumn{1}{c|}{\begin{tabular}[c]{@{}c@{}}2.667\\ 4.0\end{tabular}}    & \multicolumn{1}{c|}{\begin{tabular}[c]{@{}c@{}}4.794\\ 5.996\end{tabular}}  \\ \hline
\multicolumn{1}{|c|}{T$_\mathrm{gap}$}                                             & \multicolumn{1}{c|}{\begin{tabular}[c]{@{}c@{}}5.081\\ 8.382\end{tabular}} & \multicolumn{1}{c|}{\begin{tabular}[c]{@{}c@{}}3.427\\ 5.996\end{tabular}}  & \multicolumn{1}{c|}{\begin{tabular}[c]{@{}c@{}}7.977\\ 11.965\end{tabular}} \\ \hline
\end{tabular}
\end{table}

The two major peaks {seen} in Figure \ref{fig:regularly_spaced_sinus_saturated_1} are explained by $T_{gap}$ itself, i.e. at $T_{gap}$ and $\frac{T_{gap}}{2}$.
The two smaller peaks are explained by the combination period between $T_{obs}$ and $ T_{gap}$, for $m = 1$ and $m = 2$.
All these peaks are therefore only due to the windowing of the observations ({the} periodicity of the observation length and interval between two consecutive observations). The $95$\% confidence level is obviously at the value of the highest peak, since all data points are equal to one, and shuffling them randomly over the time intervals makes no difference to the LS analysis.

The third simulation (see Figure \ref{fig:regularly_spaced_sinus_diluted} left) is the same {as} the one in Figure \ref{fig:regularly_spaced_sinus}, to which a normal distribution was added to dilute the sinusoidal wave {in order to move a little closer to a real signal and away from the idealised case of the previous figure and to obtain an LS periodogram {that} is close to what we show in Section~\ref{sec:real_signal}.} The sinusoidal wave is still {centred} on zero, with an amplitude between $-1$ and $+1$, {and a rms of} $\sigma_\mathrm{sine} = 0.71$. The normal distribution generated has a standard deviation of $\sigma_\mathrm{normal~distribution} = K \times \sigma_\mathrm{sine}$, with $\mathrm{K}$ being the dilution factor (i.e. the signal-to-noise ratio). This distribution is then added to the sinusoidal wave, and the standard deviation of this combined signal, $\sigma_\mathrm{combined}$, is calculated. To obtain a combined signal with the same properties as the original signal, in order to decrease its signal-to-noise ratio (S/N), the combined signal is normalised by dividing by the dilution factor $K$. This new signal is called the `corrected combined signal'. Finally, as for the previous simulations, $2.65$\% of the signal is kept and gathered in $125$~interval{s} of  eight hours separated by  $\mathrm{N} \times 23.93$~hours (with $\mathrm{N} = 1, 2, 3, 4,...$) and a time resolution of $600$~seconds. 

Figure \ref{fig:regularly_spaced_sinus_diluted} shows the results with a dilution factor of $\mathrm{K}= 15$ (light grey). As one can see (left panel), the standard deviation of the original sinusoidal wave (light blue) and of the corrected combined (red) signal is the same. However, the LS periodogram (right panel) is different from that in Figure \ref{fig:regularly_spaced_sinus}: {as soon as noise is added, a forest of peaks appears.} The period of the input sine, T$_\mathrm{sine} = 12.90$~h, is still detected, even {though} the noise in the LS periodogram has drastically {increased}, and the value of the peaks has drastically {decreased}. The other two highest peaks are still located at $8.38$~hours (combination period between $T_\mathrm{gap}$ and $T_\mathrm{sine}$) and at $27.99$~hours ({beat} period between $T_\mathrm{gap}$ and $T_\mathrm{sine}$). This latter LS periodogram strongly resembles what is observed for real data, as presented in the following section. The $95$\% confidence level is located at higher LS power values, confirming that the noise is much more present in that case; however, the three highest peaks, related to $T_\mathrm{sine}$ are still well above the $95$\% level. {In this case, the noise is entirely random. Although many periods are visible in the LS periodogram (the forest of peaks), all of them remain below the $95$\% confidence level. As we show later, this will not be the case with real noise.}

{Finally, the evolution of the LS periodogram as a function of the number of samples included is presented in Appendix \ref{sec:Appendix_SNR_LS_versus_sample}, Figure \ref{fig:Figure_SNR_LS_over_samples_simulation}. For the simulated signal, the S/N of the LS periodogram -denoted S/N(LS) and calculated as the LS power divided by the standard deviation of the LS periodograms-- is evolving as the square root of the number of samples included in the LS analysis (see bottom panel). The top panel shows that for this simulated signal (diluted into noise by a factor $K = 8$) it takes $N \simeq 20$ samples for the maximal peak in the LS periodogram to be stabilised along the value of $T_\mathrm{sine}$.}

\section{Application to real signal: Jupiter observation with NenuFAR}
\label{sec:real_signal}

   In this section, we present an analysis of real observations of Jupiter's radio emissions collected using the NenuFAR radio telescope\footnote{\url{https://nenufar.obs-nancay.fr}} \citep{Zarka_2020_NenuFAR}. The Jupiter observations consist of time-frequency arrays of the Stokes parameters, which were acquired with the NenuFAR beam-forming mode throughout a dedicated long-term key programme from September 2019 to November 2022 (early science phase) and then to 2025 (regular cycles 1 to 5). The number of available mini-arrays (MA; sub--groups of $19$~antennas) increases over time (from 56 in 2019 to $\sim 80$ after April 2022, minus the ones {under} maintenance). 
   The observations used a typical $84$~ms~$\times$~$12$~kHz sampling and were scheduled close to the perijoves of the Juno spacecraft orbiting Jupiter \citep{2017SSRv..213....5B}. From cycles 2 to 4, the observations were {additionally} scheduled to search for decametric emissions induced by Io, Europa, and Ganymede \citep{2023pre9.conf03097L}. It is also worth mentioning that the effective area of NenuFAR antennas, and therefore the final instrumental sensitivity, regularly increased from 2019 to 2025.
   Overall, this large NenuFAR Jupiter dataset consists of $176$ independent observations of $\sim 8$~hours, corresponding to a total exposure of $\sim1400$~hours. The interval between the start times of two consecutive observations is $N \times 23.93$, with $N = 1, 2, 3, 4, 5, ...$, corresponding to $N \times 23$~hours, $55$~minutes, and $48$~seconds --the duration for Jupiter to return to the meridian for a fixed observer on Earth. This periodicity is also very close to the {sidereal} day, i.e. $23$~hours, $56$~minutes, $4$~seconds $\simeq 23.934$~hours. Therefore, these two periodicities will probably merge in the LS analysis, and it is hereafter referred to as $T_\mathrm{Day}$.
   
   \begin{figure}[h]
      \centering
      \includegraphics[width=1.1\columnwidth]{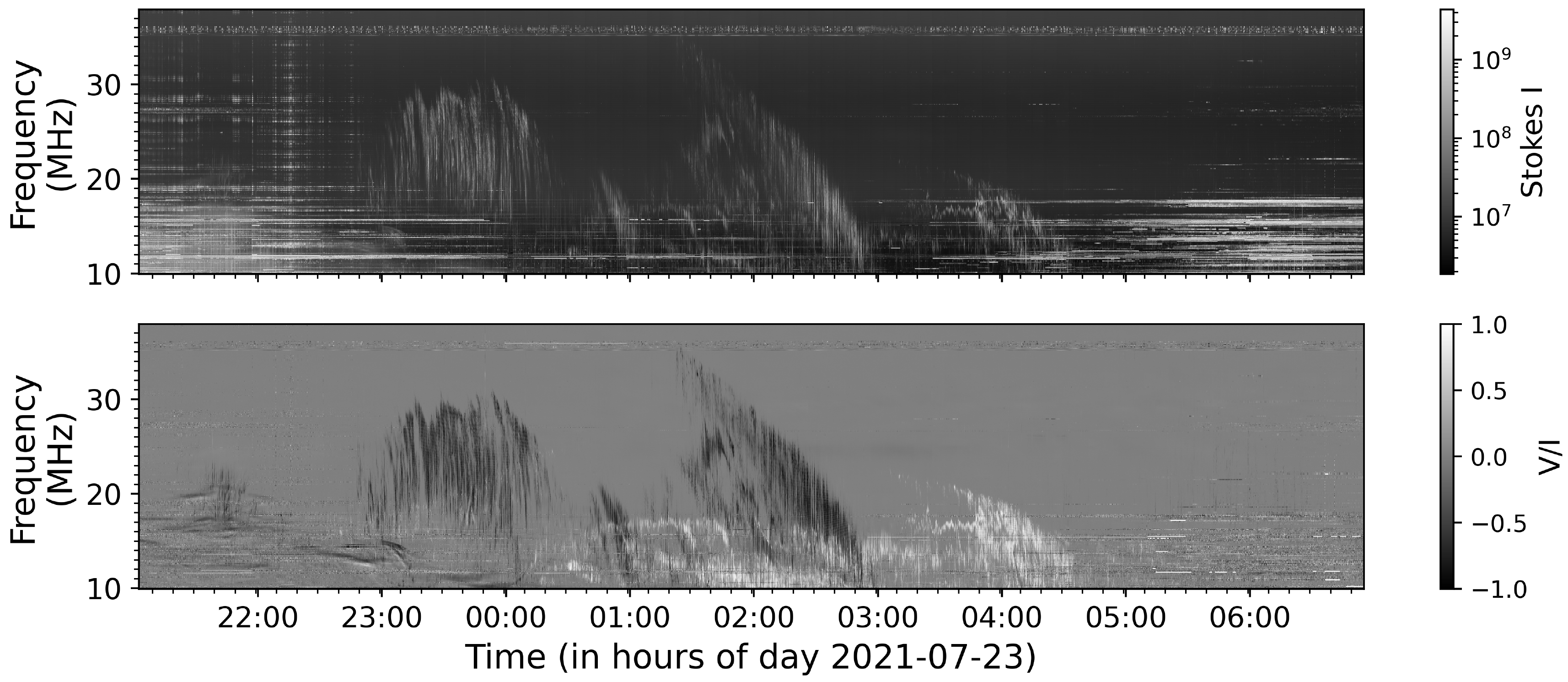}
      \caption{Typical NenuFAR `time (UTC) versus frequency (in MHz) {spectrogram}' of Jupiter signal. (Top panel) Stokes I with data preprocessing applied (see text). (Bottom panel) Ratio between Stokes $V$ and Stokes $I$ showing the degree of circular polarisation.}
      \label{fig:dynamic_spectrum}
   \end{figure}

   {A} typical observation is displayed in Figure \ref{fig:dynamic_spectrum}. {The} top panel shows the Stokes I parameter with pre-processing applied. This preprocessing of the data consists of a time--frequency integration of the data to $\sim~251$~msec~$\times \sim~24$~kHz. During this integration, {the PATROL algorithm \citep{vasylieva:tel-01246634, 2013MNRAS.431.3624Z, 2017pre8.conf..301T} is used to store} bad pixels as a weight mask of the same size (nt,nf) as the reduced Stokes I data. Each value of the flag mask is the fraction (0...1) of the good pixels integrated to obtain the corresponding pixel in the reduced data. This weight array is thresholded at 50\% (i.e. values of $\le0.5$ are set to zero and values $>0.5$ are set to one). {The} bottom panel displays the degree of circular polarisation, defined as the ratio between Stokes $V$ and Stokes $I$, with values ranging between $-1$ and $1$.

   In this figure, radio frequency interference (RFI) can clearly be seen in the Stokes I data below $20$~MHz, before 23:00 UTC and after 04:00 UTC, i.e. {during the daytime}. As these RFIs are almost not circularly polarised, they are not visible in the V/I data. In both the Stokes I and V/I ratio, radio emissions are clearly visible between 23:00 and 05:00. These emissions display different {degrees} of circular polarisation, providing insight into the polarisation characteristics of the emissions and their hemisphere of origin: V/I$>0$ corresponds to LH emissions therefore originating from the southern hemisphere of Jupiter, while V/I$<0$ corresponds to RH emissions therefore originating from the northern hemisphere of Jupiter. Using the \href{https://jupiter-probability-tool.obspm.fr/}{Jupiter Probability Tool} \citep{10.3389/fspas.2023.1091967}, these features are identified as Io-induced emissions.

   \begin{table}
    \centering
    \caption{Possible fundamental, beat, and combination periods that could be expected due to the rotation periods of Jupiter and its moons, and the day/observation periodicity}
    \label{tab:possible_periodities_NenuFAR_data}
         \small
\begin{tabular}{c|ccccc}
\cline{2-6}
                                                                                   & \multicolumn{1}{c|}{T$_\mathrm{Jupiter}$}                                   & \multicolumn{1}{c|}{T$_\mathrm{Day}$}                                       & \multicolumn{1}{c|}{T$_\mathrm{Io}$}                                        & \multicolumn{1}{c|}{T$_\mathrm{Europa}$}                                     & \multicolumn{1}{c|}{T$_\mathrm{Ganymede}$}                                   \\ \hline
\multicolumn{1}{|c|}{\begin{tabular}[c]{@{}c@{}}Period\\ Half Period\end{tabular}} & \multicolumn{1}{c|}{\begin{tabular}[c]{@{}c@{}}9.93\\ 4.97\end{tabular}}    & \multicolumn{1}{c|}{\begin{tabular}[c]{@{}c@{}}23.93\\ 11.97\end{tabular}}  & \multicolumn{1}{c|}{\begin{tabular}[c]{@{}c@{}}42.46\\ 21.23\end{tabular}}  & \multicolumn{1}{c|}{\begin{tabular}[c]{@{}c@{}}85.23\\ 42.62\end{tabular}}   & \multicolumn{1}{c|}{\begin{tabular}[c]{@{}c@{}}171.71\\ 85.855\end{tabular}} \\ \hline
\multicolumn{1}{|c|}{\begin{tabular}[c]{@{}c@{}}Beat\\ periods\end{tabular}}       &                                                                             &                                                                             &                                                                             &                                                                              &                                                                              \\ \hline
\multicolumn{1}{|c|}{T$_\mathrm{Jupiter}$}                                         & \multicolumn{1}{c|}{\begin{tabular}[c]{@{}c@{}}N/A\\ 9.93\end{tabular}}    & \multicolumn{1}{c|}{\begin{tabular}[c]{@{}c@{}}16.97\\ 58.39\end{tabular}}  & \multicolumn{1}{c|}{\begin{tabular}[c]{@{}c@{}}12.96\\ 18.66\end{tabular}}  & \multicolumn{1}{c|}{\begin{tabular}[c]{@{}c@{}}11.24\\ 12.95\end{tabular}}   & \multicolumn{1}{c|}{\begin{tabular}[c]{@{}c@{}}10.54\\ 11.23\end{tabular}}   \\ \hline
\multicolumn{1}{|c|}{T$_\mathrm{Day}$}                                             & \multicolumn{1}{c|}{\begin{tabular}[c]{@{}c@{}}16.97\\ 6.265\end{tabular}}  & \multicolumn{1}{c|}{\begin{tabular}[c]{@{}c@{}}N/A\\ 23.93\end{tabular}}    & \multicolumn{1}{c|}{\begin{tabular}[c]{@{}c@{}}54.83\\ 188.2\end{tabular}}  & \multicolumn{1}{c|}{\begin{tabular}[c]{@{}c@{}}33.27\\ 54.58\end{tabular}}   & \multicolumn{1}{c|}{\begin{tabular}[c]{@{}c@{}}27.81\\ 33.18\end{tabular}}   \\ \hline
\multicolumn{1}{|c|}{T$_\mathrm{Io}$}                                              & \multicolumn{1}{c|}{\begin{tabular}[c]{@{}c@{}}12.96\\ 5.623\end{tabular}}  & \multicolumn{1}{c|}{\begin{tabular}[c]{@{}c@{}}54.83\\ 16.66\end{tabular}}  & \multicolumn{1}{c|}{\begin{tabular}[c]{@{}c@{}}N/A\\ 42.46\end{tabular}}    & \multicolumn{1}{c|}{\begin{tabular}[c]{@{}c@{}}84.61\\ 11673.8\end{tabular}} & \multicolumn{1}{c|}{\begin{tabular}[c]{@{}c@{}}56.41\\ 84.01\end{tabular}}   \\ \hline
\multicolumn{1}{|c|}{T$_\mathrm{Europa}$}                                          & \multicolumn{1}{c|}{\begin{tabular}[c]{@{}c@{}}11.24\\ 5.272\end{tabular}}  & \multicolumn{1}{c|}{\begin{tabular}[c]{@{}c@{}}33.27\\ 13.92\end{tabular}}  & \multicolumn{1}{c|}{\begin{tabular}[c]{@{}c@{}}84.61\\ 28.27\end{tabular}}  & \multicolumn{1}{c|}{\begin{tabular}[c]{@{}c@{}}N/A\\ 85.23\end{tabular}}     & \multicolumn{1}{c|}{\begin{tabular}[c]{@{}c@{}}169.2\\ 11707.9\end{tabular}} \\ \hline
\multicolumn{1}{|c|}{T$_\mathrm{Ganymede}$}                                        & \multicolumn{1}{c|}{\begin{tabular}[c]{@{}c@{}}10.54\\ 5.113\end{tabular}}  & \multicolumn{1}{c|}{\begin{tabular}[c]{@{}c@{}}27.81\\ 12.86\end{tabular}}  & \multicolumn{1}{c|}{\begin{tabular}[c]{@{}c@{}}56.41\\ 24.23\end{tabular}}  & \multicolumn{1}{c|}{\begin{tabular}[c]{@{}c@{}}169.2\\ 56.68\end{tabular}}   & \multicolumn{1}{c|}{\begin{tabular}[c]{@{}c@{}}N/A\\ 171.71\end{tabular}}    \\ \hline
\multicolumn{1}{|c|}{\begin{tabular}[c]{@{}c@{}}combination\\ periods\end{tabular}}   &                                                                             &                                                                             &                                                                             &                                                                              &                                                                              \\ \cline{1-1}
\multicolumn{1}{|c|}{n = 1}                                                        & \multicolumn{1}{l}{}                                                        & \multicolumn{1}{l}{}                                                        & \multicolumn{1}{l}{}                                                        & \multicolumn{1}{l}{}                                                         & \multicolumn{1}{l}{}                                                         \\ \hline
\multicolumn{1}{|c|}{T$_\mathrm{Jupiter}$}                                         & \multicolumn{1}{c|}{\begin{tabular}[c]{@{}c@{}}4.965\\ 6.620\end{tabular}}  & \multicolumn{1}{c|}{\begin{tabular}[c]{@{}c@{}}7.018\\ 8.224\end{tabular}}  & \multicolumn{1}{c|}{\begin{tabular}[c]{@{}c@{}}8.048\\ 8.890\end{tabular}}  & \multicolumn{1}{c|}{\begin{tabular}[c]{@{}c@{}}8.894\\ 9.383\end{tabular}}   & \multicolumn{1}{c|}{\begin{tabular}[c]{@{}c@{}}9.387\\ 9.651\end{tabular}}   \\ \hline
\multicolumn{1}{|c|}{T$_\mathrm{Day}$}                                             & \multicolumn{1}{c|}{\begin{tabular}[c]{@{}c@{}}7.018\\ 10.853\end{tabular}} & \multicolumn{1}{c|}{\begin{tabular}[c]{@{}c@{}}11.965\\ 15.95\end{tabular}} & \multicolumn{1}{c|}{\begin{tabular}[c]{@{}c@{}}15.31\\ 18.67\end{tabular}}  & \multicolumn{1}{c|}{\begin{tabular}[c]{@{}c@{}}18.69\\ 20.99\end{tabular}}   & \multicolumn{1}{c|}{\begin{tabular}[c]{@{}c@{}}21.00\\ 22.37\end{tabular}}   \\ \hline
\multicolumn{1}{|c|}{T$_\mathrm{Io}$}                                              & \multicolumn{1}{c|}{\begin{tabular}[c]{@{}c@{}}8.048\\ 13.53\end{tabular}}  & \multicolumn{1}{c|}{\begin{tabular}[c]{@{}c@{}}15.31\\ 22.50\end{tabular}}  & \multicolumn{1}{c|}{\begin{tabular}[c]{@{}c@{}}21.23\\ 28.31\end{tabular}}  & \multicolumn{1}{c|}{\begin{tabular}[c]{@{}c@{}}28.34\\ 33.99\end{tabular}}   & \multicolumn{1}{c|}{\begin{tabular}[c]{@{}c@{}}34.04\\ 37.79\end{tabular}}   \\ \hline
\multicolumn{1}{|c|}{T$_\mathrm{Europa}$}                                          & \multicolumn{1}{c|}{\begin{tabular}[c]{@{}c@{}}8.894\\ 16.11\end{tabular}}  & \multicolumn{1}{c|}{\begin{tabular}[c]{@{}c@{}}18.69\\ 30.65\end{tabular}}  & \multicolumn{1}{c|}{\begin{tabular}[c]{@{}c@{}}28.34\\ 42.54\end{tabular}}  & \multicolumn{1}{c|}{\begin{tabular}[c]{@{}c@{}}42.62\\ 56.82\end{tabular}}   & \multicolumn{1}{c|}{\begin{tabular}[c]{@{}c@{}}56.96\\ 68.28\end{tabular}}   \\ \hline
\multicolumn{1}{|c|}{T$_\mathrm{Ganymede}$}                                        & \multicolumn{1}{c|}{\begin{tabular}[c]{@{}c@{}}9.387\\ 17.80\end{tabular}}  & \multicolumn{1}{c|}{\begin{tabular}[c]{@{}c@{}}21.00\\ 37.44\end{tabular}}  & \multicolumn{1}{c|}{\begin{tabular}[c]{@{}c@{}}34.04\\ 56.82\end{tabular}}  & \multicolumn{1}{c|}{\begin{tabular}[c]{@{}c@{}}56.96\\ 85.54\end{tabular}}   & \multicolumn{1}{c|}{\begin{tabular}[c]{@{}c@{}}85.86\\ 114.5\end{tabular}}   \\ \hline
\multicolumn{1}{|c|}{n=2}                                                          &                                                                             &                                                                             &                                                                             &                                                                              &                                                                              \\ \hline
\multicolumn{1}{|c|}{T$_\mathrm{Jupiter}$}                                         & \multicolumn{1}{c|}{\begin{tabular}[c]{@{}c@{}}3.310\\ 4.965\end{tabular}}  & \multicolumn{1}{c|}{\begin{tabular}[c]{@{}c@{}}5.427\\ 7.0179\end{tabular}} & \multicolumn{1}{c|}{\begin{tabular}[c]{@{}c@{}}6.766\\ 8.048\end{tabular}}  & \multicolumn{1}{c|}{\begin{tabular}[c]{@{}c@{}}8.053\\ 8.894\end{tabular}}   & \multicolumn{1}{c|}{\begin{tabular}[c]{@{}c@{}}8.901\\ 9.387\end{tabular}}   \\ \hline
\multicolumn{1}{|c|}{T$_\mathrm{Day}$}                                             & \multicolumn{1}{c|}{\begin{tabular}[c]{@{}c@{}}4.112\\ 7.018\end{tabular}}  & \multicolumn{1}{c|}{\begin{tabular}[c]{@{}c@{}}7.977\\ 11.97\end{tabular}}  & \multicolumn{1}{c|}{\begin{tabular}[c]{@{}c@{}}11.25\\ 15.31\end{tabular}}  & \multicolumn{1}{c|}{\begin{tabular}[c]{@{}c@{}}15.33\\ 18.68\end{tabular}}   & \multicolumn{1}{c|}{\begin{tabular}[c]{@{}c@{}}18.71\\ 21.00\end{tabular}}   \\ \hline
\multicolumn{1}{|c|}{T$_\mathrm{Io}$}                                              & \multicolumn{1}{c|}{\begin{tabular}[c]{@{}c@{}}4.445\\ 8.048\end{tabular}}  & \multicolumn{1}{c|}{\begin{tabular}[c]{@{}c@{}}9.335\\ 15.31\end{tabular}}  & \multicolumn{1}{c|}{\begin{tabular}[c]{@{}c@{}}14.15\\ 21.23\end{tabular}}  & \multicolumn{1}{c|}{\begin{tabular}[c]{@{}c@{}}21.27\\ 28.34\end{tabular}}   & \multicolumn{1}{c|}{\begin{tabular}[c]{@{}c@{}}28.41\\ 34.04\end{tabular}}   \\ \hline
\multicolumn{1}{|c|}{T$_\mathrm{Europa}$}                                          & \multicolumn{1}{c|}{\begin{tabular}[c]{@{}c@{}}4.692\\ 8.894\end{tabular}}  & \multicolumn{1}{c|}{\begin{tabular}[c]{@{}c@{}}10.49\\ 18.68\end{tabular}}  & \multicolumn{1}{c|}{\begin{tabular}[c]{@{}c@{}}16.996\\ 28.34\end{tabular}} & \multicolumn{1}{c|}{\begin{tabular}[c]{@{}c@{}}28.41\\ 42.62\end{tabular}}   & \multicolumn{1}{c|}{\begin{tabular}[c]{@{}c@{}}42.77\\ 56.96\end{tabular}}   \\ \hline
\multicolumn{1}{|c|}{T$_\mathrm{Ganymede}$}                                        & \multicolumn{1}{c|}{\begin{tabular}[c]{@{}c@{}}4.825\\ 9.387\end{tabular}}  & \multicolumn{1}{c|}{\begin{tabular}[c]{@{}c@{}}11.19\\ 21.00\end{tabular}}  & \multicolumn{1}{c|}{\begin{tabular}[c]{@{}c@{}}18.894\\ 34.04\end{tabular}} & \multicolumn{1}{c|}{\begin{tabular}[c]{@{}c@{}}34.14\\ 56.96\end{tabular}}   & \multicolumn{1}{c|}{\begin{tabular}[c]{@{}c@{}}57.24\\ 85.86\end{tabular}}   \\ \hline
\end{tabular}
\end{table}

   In the following, the focus is on this degree  of circular polarisation, since RFI and sky background have little or no circular polarisation, thereby resulting in a better S/N. {In addition, the fact that $|V/I| \leq 1$ --while the values of $|V|$ vary over several orders of magnitude-- allows the LS method to work better.}

   Using the \cite{louis_2025_15078360} pipeline, the observations were  further {processed} to a time resolution of 600 seconds and a frequency resolution of 1 MHz, ranging from 8~MHz to 88~MHz, to further increase the S/N \citep[to access the processed data, please see][]{2025NenuFAR_KP07_Jupiter_dataset_preprocessed_LS}.
   Note here that the data up to 88~MHz were kept, since they were acquired up to this frequency from April 2023, with a view to observing synchrotron radiation. However, the circular polarisation expected for this radiation is weak-to-non-detectable in this frequency range \citep{2016A&A...587A...3G}, as Jupiter's radiation belt system is not resolved with NenuFAR resolution, which may lead to circular polarisation smearing.
   {Therefore}, only circular polarisation data up to 40 MHz are analysed, i.e. the upper limit of auroral radiation.

{Table} \ref{tab:possible_periodities_NenuFAR_data} lists the {fundamental and periods at which one can expect to see signals for Jupiter, $T_\mathrm{Jupiter}$ (corresponding to auroral radio emissions), and its various moons {that} control part of the Jovian radio spectrum: Io, $T_\mathrm{Io}$; Europa, $T_\mathrm{Europa}$; and Ganymede, $T_\mathrm{Ganymede}${.} In addition,  the day/observation periodicity, $T_\mathrm{Day}$, was added. The different combinations of these periods are listed; i.e. the beat and combination periods that can be expected to be found in the LS periodograms that follow.} {The average length of the observations are not included in the calculation of the beat or {combination} periods; as seen in Section \ref{sec:real_signal}, this does not influence the periodograms (except in the case of the study of pure windowing).}

\begin{figure}[]
   \centering
   \includegraphics[width=1\columnwidth]{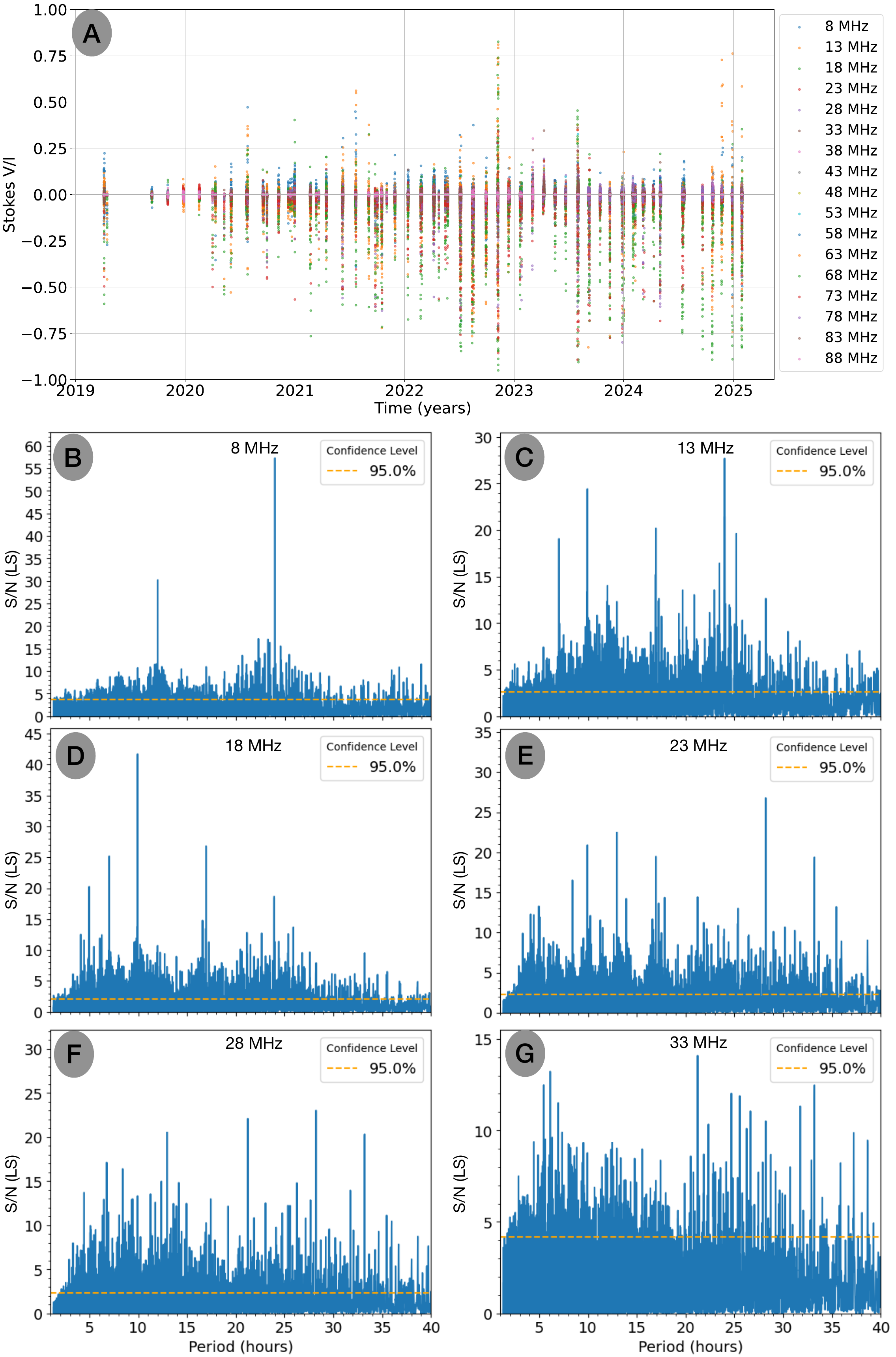}
   \caption{(A) Real data from NenuFAR observations. The colour corresponds to different frequency ranges of 5 MHz. (B--G) Corresponding periodograms of the S/N of the LS power (noted S/N(LS))for seven frequency ranges averaged on a $5$~MHz bandwidth: (B) [$8${, }$13$[~MHz, (C) [$13${, }$18$[~MHz, (D) [$18${, }$23$[~MHz, (E) [$23${, }$28$[~MHz, (F) [$28${, }$33$[~MHz, (G) [$33${, }$38$[~MHz. The {$95$\% confidence level is also shown in the LS periodograms.}}
   \label{fig:NenuFAR_data_all_freq_periodogram}
\end{figure}

Figure \ref{fig:NenuFAR_data_all_freq_periodogram}A shows the degree of circular polarisation (Stokes V/I) measurements within 5~MHz frequency bands. Figure{s} \ref{fig:NenuFAR_data_all_freq_periodogram}B--G show periodograms of the {S/N} of the LS power --noted S/N(LS)-- for 5~MHz frequency bands in the [$8${, }$38$[~MHz range. The {level of the} noise in each frequency band is approximated by the standard deviation of the LS periodogram. {The $95$\% confidence level is shown in each panel. In the case of real data, this confidence level is still quite low, and well below a large number of peaks. As indicated at the end of the Section \ref{sec:simu}, for data with `structured windows', this produces data aliasing and high noise correlation. As a result, the FAP tends to be underestimated by the bootstrap method} {\citep[see, e.g. Section 7.4.2.3 in][]{2018ApJS..236...16V}}.
{To partially eliminate noise and enable comparison between different frequency bands (where noise levels may vary), the LS periodogram is plotted as a function of the S/N, calculated as the LS power (in each frequency band) divided  by  the  standard  deviation  of  the  LS  periodograms (in each frequency band). It should be noted that this S/N should not be considered as a confidence sigma, but rather it serves to define a common threshold for all frequency bands. Figure~\ref{fig:2D_periodogram}} shows a summary in 2D (observed radio frequency versus LS periodicities) of Figs. \ref{fig:NenuFAR_data_all_freq_periodogram}B--G for $1$~MHz bands. {The data in this figure were thresholded by calculating the 99.99th percentile of the overall 2D distribution, resulting in a value of S/N(LS) = $15$.}

\begin{figure}[]
   \centering
   \includegraphics[width=1\columnwidth]{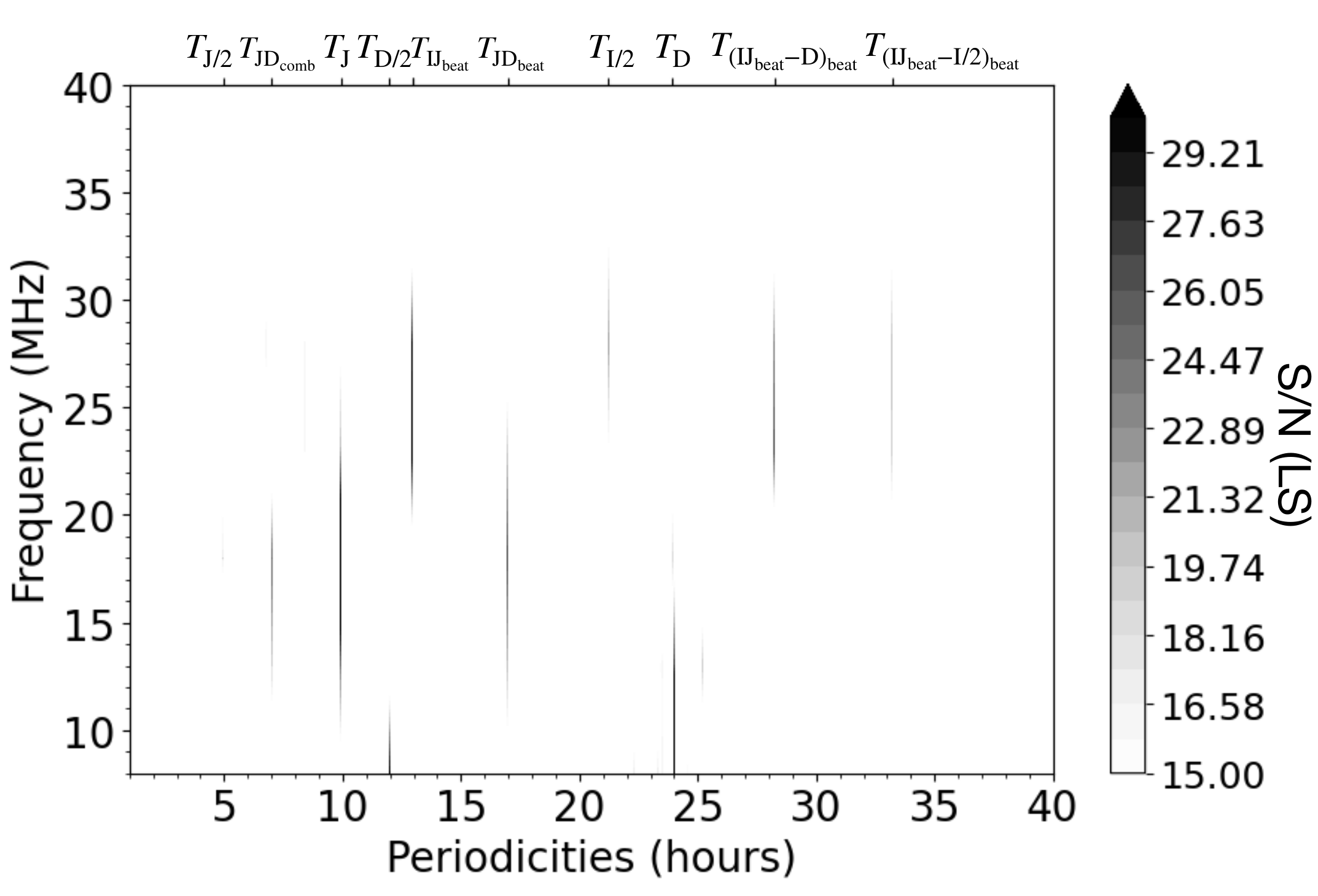}
\caption{S/N of 2D LS periodogram. {The} Y-axis {represents} the observed frequencies (from 8 to 43 MHz), and the X-axis represents the LS periodicities. Periodicities of interest are indicated above the top Y-axis.}
 \label{fig:2D_periodogram}
\end{figure}

Depending on the frequency, different peaks are visible. First, peaks are detected at $9.93$~hours within the range of $\sim13$--$25$~MHz and at $4.97$~hours between $19$--$24$~MHz, which corresponds to $T_\mathrm{J}$, the fundamental of the Jupiter rotation period, and $T_\mathrm{J/2,}$ half a Jupiter rotation period. Two other peaks are linked to Io: one detected at $12.96$~hours from $\sim 23$--$33$~MHz, which correspond{s} to $T_\mathrm{{beat}~IJ,}$ the {beat} period between Jupiter and its Galilean moon Io (which corresponds to the synodic period of Io); and a second one at $21.23$~hours from $26--33$~MHz, corresponding to $T_\mathrm{Io/2,}$ which is Io's half orbital period, implying that the LS analysis detects radio emissions related to Jupiter itself and induced by Io (at Io's revolution or Io--Jupiter {beat,} or in that case synodic periods). {The higher S/N(LS) for $T_{\mathrm{Jupiter-Io}_\mathrm{beat}}$ than for $T_\mathrm{Io}$ highlights the fact that Jupiter's magnetic field is not axisymmetric. This is furthermore {emphasised} by the S/N(LS) values that are higher for $T_\mathrm{Io/2}$ than for $T_\mathrm{Io}$.}
Peaks related to Jupiter are seen at lower frequencies than {those} related to Io, which is due to the topology of the Jovian magnetic fields \citep[Io-induced radio emissions reach higher {frequencies}, see, e.g., ][]{2017A&A...604A..17M}.

Below $\sim 25$~MHz, {another} strong peak is detected at $\sim 23.93$~hours. This corresponds to $T_\mathrm{day}$, which is both the sidereal period for Jupiter to be visible in the sky and the sidereal day. The former periodicity only affects the Jovian auroral radio emissions (usually seen at $< 25$~MHz, due to Jupiter magnetic-field topology connected to the magnetic-field lines producing these emissions) as they can be observed almost every time Jupiter is visible in the sky, while Io-induced emission{s} are visible only if Io is in quadrature. The latter affects the background radio signal visibility --the RFI is most probably still visible in the V/I data-- which {is} also only visible at $\le 20$~MHz (see Figure \ref{fig:dynamic_spectrum}), since this periodicity is only detected at low {frequencies}. {A} harmonic of this peak is also visible at T$_\mathrm{D/2}$.

Four other peaks are also visible. Two above $23$~{MHz} are related to Io's periods (Keplerian and {beat} with Jupiter): (i) at $28.27$~hours from $23$--$33$~MHz  corresponding to $T_\mathrm{{beat}~D-IJ,}$ which is the {beat} period between the $T_\mathrm{{beat}~IJ}$ and $T_\mathrm{D}$ periodicities; (ii) at $33.27$~hours from $24$--$32$~MHz, corresponding to $T_\mathrm{{beat}~I/2-IJ,}$ which is the {beat} period between $T_\mathrm{Io/2,}$ i.e. the Io's half-orbital period, and the $T_\mathrm{{beat}~IJ}$ period. Two are contained in the $13$--$25$~MHz range and are related to Jupiter's period: (i) T$_\mathrm{comb~JD,}$ which is the combination period between the Jupiter, $T_\mathrm{J,}$ and the day, $T_\mathrm{D,}$ periodicities at $7.02$~hours from $13$--$23$~MHz and (ii) $T_\mathrm{{beat}~DJ,}$ which is the {beat} period between the Jupiter, $T_\mathrm{J,}$ and the day, $T_\mathrm{D,}$ periodicities at $16.97$~hours from $13$--$25$~MHz.

\begin{figure*}[]
   \includegraphics[width=1\textwidth]{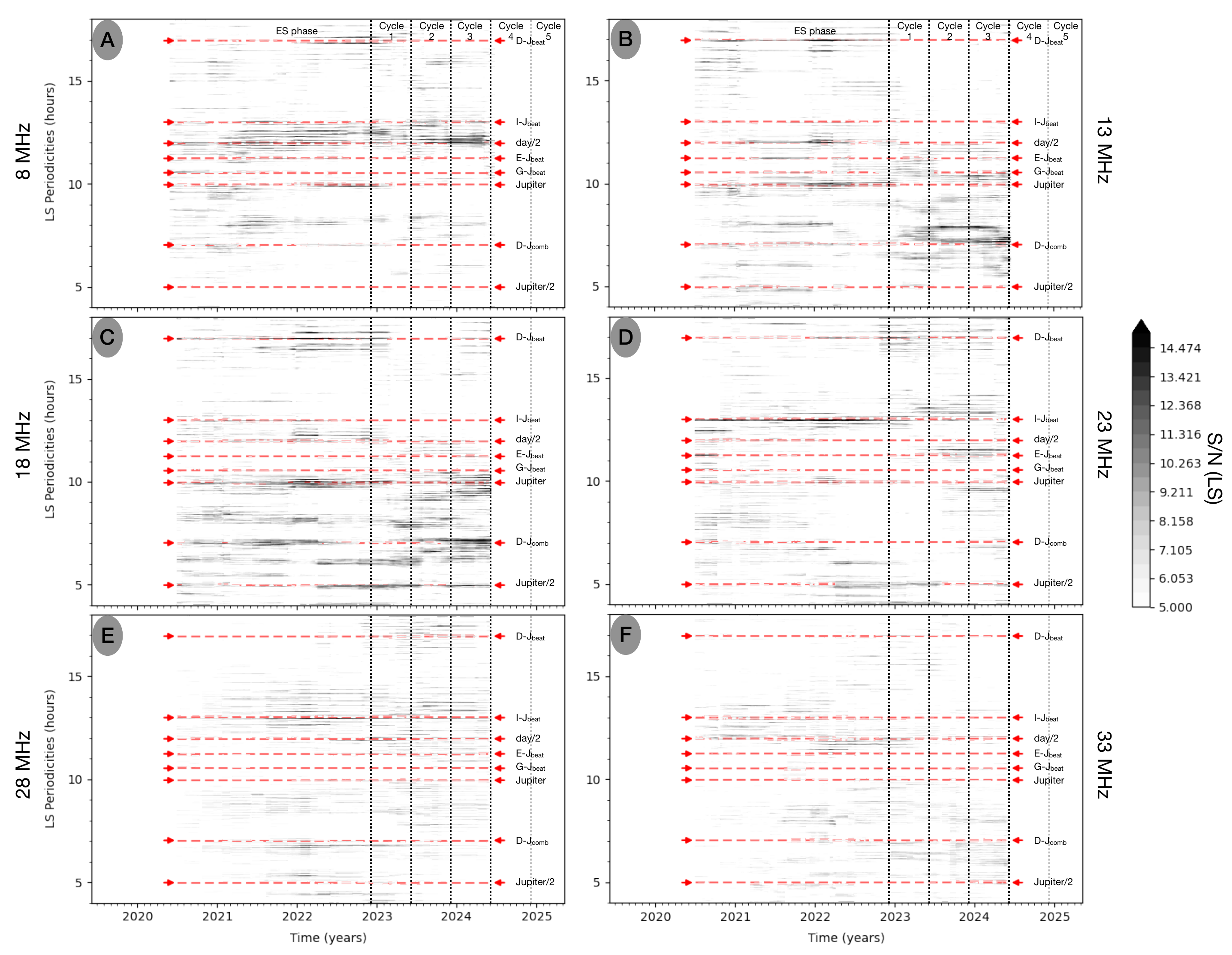}
\caption{S/N of 2D LS periodograms for six different 5-MHz frequency intervals ([$8${, }$13$[~MHz, [$13${, }$18$[~MHz, [$18${, }$23$[~MHz, [$23${, }$28$[~MHz, [$28${, }$33$[~MHz, and [$33${, }$38$[~MHz). In each panel, the Y-axis represents the LS periodicities, and the X-axis represents the calendar time (in {years}).  The LS periodograms are calculated over a $500$~day sliding window, which shifts by two days. The mean time is taken for each window, and the corresponding S/N of the LS periodogram is displayed in grey. The colour bar, thresholded to remove background noise from the periodograms, is the same for all panels. The different dashed red lines give (from top to bottom) the following periodicities: $T_\mathrm{{beat}~D-J}$, $T_\mathrm{{beat}~I-J}$, $T_\mathrm{day/2}$, $T_\mathrm{{beat}~E-J}$ (for Europa--Jupiter), $T_\mathrm{{beat}~G-J}$ (for Ganymede--Jupiter), $T_\mathrm{Jupiter}$, $T_\mathrm{comb~D-J}$, and $T_\mathrm{Jupiter/2}$. The vertical dotted line {indicates} the NenuFAR observing phase, from Early Science to Cycle 5.
}
 \label{fig:LS_over_time}
\end{figure*}

Interestingly, the peaks at T$_\mathrm{comb~DJ}$, $T_\mathrm{{beat}~DJ}$, $T_\mathrm{{beat}~D-IJ,}$ and $T_\mathrm{{beat}~I/2-IJ}$ each {combine} individual LS power at two different frequencies, {thereby} strengthening the detection of periodic signals at $T_\mathrm{{beat}~IJ}$, $T_\mathrm{I/2,}$ and $T_\mathrm{J}$, providing {stronger evidence} for these periodicities. {Figure \ref{fig:Figure_SNR_LS_over_samples_data} displays (bottom panel) the evolution of the S/N(LS) as a function of observation samples inserted into the LS periodogram for some of the major LS periodicity peaks. It shows that some of beat or combination periodicity peaks (e.g. T$_\mathrm{comb~DJ}$ and $T_\mathrm{{beat}~DJ}$) can be almost as high as fundamental periodicities (e.g., $T_\mathrm{{beat}~IJ}$ or $T_\mathrm{Io}$). Figure \ref{fig:Figure_SNR_LS_over_samples_simulation}, top panel, shows that it takes $ N \gtrsim 4$--$15$ samples (i.e. observations) to achieve stability in the values of the periodicities.}

Other potential periodicities could be linked to two additional Galilean moons --Europa and Ganymede-- as they also induce radio emissions in the decametric range. \citep{2017JGRA..122.9228L, 2023JGRA..12831985L, 2018A&A...618A..84Z, 2022A&A...665A..67J}. As for Io, the expected periodicity would have been at the {beat} period with Jupiter, i.e. $\mathrm{T}_\mathrm{{beat}~Europa-Jupiter} = 11.24$~hours and $\mathrm{T}_\mathrm{{beat}~Ganymede-Jupiter} =10.54$~hours. No significant peaks are visible at these periodicities; only a small one is observed at $\mathrm{T}_\mathrm{{beat}~Europa-Jupiter}$ (LS power = $\sim 0.014$) in Figures \ref{fig:NenuFAR_data_all_freq_periodogram}E--F. This is not a surprise, as almost all radio {emissions} detected in the NenuFAR Jupiter observation {are} known to be associated with the Io--Jupiter interaction or to Jupiter itself. The periodic signals at  $\mathrm{T}_\mathrm{{beat}~Europa-Jupiter}$ and $\mathrm{T}_\mathrm{{beat}~Ganymede-Jupiter}$ are further discussed in the following paragraphs.

Based on the detected peaks at specific periodicities, the goal is to constrain the times at which the signals contributing to the LS power were observed. The LS periodogram, initially calculated over the data acquired during an interval in excess of six years, raises the question of whether similar analyses could be performed over shorter time spans. To address this, Figure \ref{fig:LS_over_time} displays the S/N of the LS periodograms calculated over time for different frequency ranges using a sliding window approach. Each window spans $500$~days and slides every two days. Note that the effect of the window size and slide was studied: after several tests, it was determined that a window size of $200$--$500$ days and a slide of the window of $ \lesssim 1\%$ of the window size {yields} the best results. A window that is too small (e.g. $100$~days) induces a large increase {in} the LS noise. {This can also be understood by looking at Figure \ref{fig:Figure_SNR_LS_over_samples_data}: 200 days corresponds to approximately 20 samples.}

Figure \ref{fig:LS_over_time} is a {zoomed-in view of} the $4$--$20$~hour LS periodicities highlighting most of the periodicities of interest, including $T_\mathrm{{beat}~D-J}$, $T_\mathrm{{beat}~I-J}$, $T_\mathrm{day/2}$, $T_\mathrm{{beat}~E-J}$ (for Europa--Jupiter), $T_\mathrm{{beat}~G-J}$ (for Ganymede--Jupiter), $T_\mathrm{Jupiter}$, $T_\mathrm{comb~D-J}$, and $T_\mathrm{Jupiter/2}$, presented from top to bottom in each panel.
For the [$8${, }$18$[~MHz range (Figure \ref{fig:LS_over_time}A), a notably large peak is {centred} at $T_\mathrm{day/2} = 11.97$~hours due to {both} the Earth rotation period {and} the time it takes for Jupiter {to return to the meridian}.
Some peaks are closely aligned with Jupiter periodicities at $T_\mathrm{Jupiter}$, $T_\mathrm{Jupiter/2,}$ or at $T_\mathrm{{beat}~D-J}$ and $T_\mathrm{comb~D-J}$; or with the Io--Jupiter {beat} period of T$_\mathrm{{beat}~I-J}$ (e.g. Figures \ref{fig:LS_over_time}D--F). Interestingly, an enhancement of the S/N of the LS power is visible at $T_\mathrm{{beat}~E-J}$ in the [$13${, }$18$[~MHz frequency range in 2021 (panel \ref{fig:LS_over_time}B) and in the [$23${, }$28$[ frequency range in 2023--2024 from mid-cycle 2 to mid-cycle 3 (panel \ref{fig:LS_over_time}D) and at $T_\mathrm{{beat}~G-J}$ in the [$13${, }$18$[~MHz frequency range mainly during cycle 1 and also from time to time during the early science phase $\leq 2022$ (panel \ref{fig:LS_over_time}B). These results emphasise the potential of LS periodograms, calculated over shorter time spans, to reveal temporal variations in signal periodicities and to identify specific intervals of enhanced observational significance.

\section{Summary and discussions}
\label{sec:summary}
In this article, we {analyse} and demonstrate how the use of the LS periodogram is affected by a sporadic signal that is more or less regular in time, and more or less diluted with noise. The different peaks in the periodogram were analysed to determine their association with real signals.

First, a sinusoidal signal was simulated with a periodicity of $T = 12.9$ hours and an amplitude between $-1$ and $1$, spanning  five years. Only 2.65\% of the data points were randomly retained. When the LS periodogram was applied, the highest peak was clearly located at 12.9 hours, corresponding to the input signal's periodicity.

In the second simulation, the retained data were grouped into 125 intervals, each lasting eight hours, separated by $N \times 23.93$ hours ($N = 1, 2, 3, \dots$). The LS periodogram still showed the highest peak at $12.9$~hours, but additional peaks appeared with significant power at periodicities such as $27.8$~hours, $8.4$~hours, and $6.2$~hours. These correspond to the {beat} or combination periods of the input signal's periodicity and the interval between observations.

In the third simulation, noise was introduced by adding a normal distribution to dilute the signal. This increased the noise in the LS periodogram, causing a drastic decrease in the normalised peak power. The periodogram started resembling those of real observed signals. Despite this, with a dilution factor of up to $15$, the input signal's periodicity remained easily detectable, standing out above the noise with a confidence level above $95$\%. Additionally, two other peaks related to the {beat} period and {combination} period between the input signal and the observation gap were observed. Even with a diluted signal, the three main peaks retained their link to the input signal, strengthening the detection.

The LS periodogram was then applied to real observations of the circular degree of polarisation of Jupiter's radio emissions, collected with the NenuFAR radio telescope and integrated over $10$~minutes and $1$~MHz (Figure \ref{fig:2D_periodogram}) or $5$~MHz (Figure \ref{fig:NenuFAR_data_all_freq_periodogram}). It should be noted that time and frequency integration are relatively important if LS analysis is to be effective; integrating over the whole frequency band greatly dilutes the signal and {decreases} the S/N, and therefore it does not allow any periodicities to be detected; keeping the initial resolution ($12$~{kHz}) does not give enough weight to the signal, which is located over a wider frequency range. It is therefore necessary to produce several data sets with different integrations. We found that integrating over $1$~MHz is optimal, giving a good S/N and greater detail in the 2D plot (see Figure \ref{fig:2D_periodogram}), but integration over $5$~MHz is sufficient to focus on the major trends (see Figure \ref{fig:NenuFAR_data_all_freq_periodogram}). Regarding the time integration, Jovian emissions studied here are often observed over several tens of minutes to several hours (see Figure \ref{fig:dynamic_spectrum}), and periodicities of the order of an hour to tens of hours were searched in the present case. Keeping the original resolution ($84$~msec) is therefore not useful and would increase the time required to calculate the LS periodogram. We determined through several runs that integration over ten minutes is the most appropriate in this case.

Depending on the analysis method --either over the entire six-year interval or using a sliding window of $500$~days with a two-day step-- and the observed frequency range, different peaks were detected. At low frequencies, the highest peak was found at $23.93$~hours, corresponding to both the {sidereal} day and the {interval between Jupiter's meridian transits}. A smaller peak was also detected at half this period. These peaks are only visible at low frequencies for two probable reasons: (i) there are still some RFIs present in the V/I data, and these RFIs are only visible below 20 MHz during the day; (ii) only the Jovian auroral radio emissions (usually seen at $<25$~MHz) affect these periodicities, as they can be observed each time Jupiter is visible in the sky, while Io-induced {emissions} are visible only if Io is in quadrature. Four peaks were detected that are associated with Jupiter's full- and half-rotation periods, the {beat} period between Jupiter and its moon Io, and half of Io's orbital period. 
Four additional peaks linked to Jupiter or Io-induced emissions were detected,
at the {beat} and combination periods between Jupiter's and Earth's day length,
between the day and the {beat} period between Jupiter and its moon Io, and between half of Io's Keplerian orbital period and the {beat} period between Jupiter and its moon Io. {In addition, using the sliding-window technique, it enabled the detection of additional peaks at the beat period between Europa and Jupiter and at the beat period between Ganymede and Jupiter for a given period of time. The non-detection of these emissions in the 2D periodogram of Figure \ref{fig:2D_periodogram} is not a surprise; these signals are very sporadic, and the NenuFAR observations were not programmed to search for these emissions in particular, they were programmed to support the Juno mission \citep{2017SSRv..213....5B}. It should be remembered that these emissions were only recently detected, either on a case-by-case basis by comparison with simulations of their occurrence \citep{2017JGRA..122.9228L} or via statistical methods based on 26-years of data from daily observations \citep{2018A&A...618A..84Z,2022A&A...665A..67J}.} Although emissions induced by these two moons are known to exist, their detection here may be questionable given the 1:2:4 resonance among Io, Europa, and Ganymede. Indeed, there is no filtering or re-organisation of the data before applying the LS analysis. Thus, Io's emissions could disrupt the LS analysis, and their one-in-two or one-in-four detection could give power to the LS Jupiter--Europa or Jupiter--Ganymede {beat} period. However, looking at panel \ref{fig:LS_over_time}B, the S/N increase at the {beat} periods of Europa and Ganymede is not observed at the same time, and little-to-no signal is detected at the {beat} period of Io at these times, reinforcing their detection.

For the sake of completeness, this technique was also tested on a dataset covering a longer period of time and, above all, pre-catalogued (see Appendix \ref{sec:Appendix_LS_NDA}). To this end, the LS analysis was performed on the Nan\c cay Decameter Array (NDA) database built by \citet{2017A&A...604A..17M}, which separates Io-induced from non-Io-induced emissions. Figure \ref{fig:Appendix_LS_NDA_SNR_2D} (top panel) shows a 2D periodogram of all the catalogued NDA data. This periodogram shows, as the one in Figure \ref{fig:2D_periodogram}, S/N enhancements at typical periods ($T_\mathrm{Jupiter}$, $T_\mathrm{{beat}~I/J}$, $T_\mathrm{Day}$, $T_\mathrm{Day/2}$, and {beat} and combination periods between these different periodicities), while showing stronger S/Ns and peaks at many other {beat} and {combination} frequencies that we do not describe in detail here. As this LS analysis was done on a large dataset ($26$~years of data with  eight hours of daily {observations}) of catalogued data, it shows an upper limit to what a LS analysis can {provide}. Finally, the middle and bottom panels of Figure \ref{fig:Appendix_LS_NDA_SNR_2D} show LS periodograms of Io and non-Io data, respectively. Comparing them confirms what was described above; i.e. that non-Io emissions (and therefore mainly Jupiter) are only observed below $\sim25$~MHz, and that the peaks at $T_\mathrm{day}$ and $T_\mathrm{day/2}$ {are} mainly due to the return of the auroral Jupiter radio emissions (i.e. non-Io induced) in the observer's sky, and not much by the Earth's rotation period (i.e. the RFI).

Concerning the synchrotron radiation, {which is briefly mentioned in Section \ref{sec:real_signal},} {to detect the weak circularly polarised signal for this radiation the resolution of the instrument must be at least equal to half the diameter of Jupiter in order to resolve the eastern and western components of the radiation belts (and to avoid smearing)}. Upcoming SKA--low stations \citep{SKA2022_report2} might {provide} the necessary baseline to resolve this emission at low frequencies (down to $50$~MHz).

These results demonstrate that the LS periodogram is a powerful tool for detecting periodic radio emissions from unevenly sampled data. It not only identifies strong signals, such as Jupiter's auroral radio emissions and Io-induced emissions, it also detects weaker or more sporadic signals such as those linked to Europa or Ganymede. 

At first glance, one might think that regular gaps in observation would weaken the analysis by introducing spurious periodic signals. {While it is true that this regularity produces a peak in the periodograms (at $T_\mathrm{day}$)}, it also creates peaks of {combination} or {beat} periods with the real signal {(e.g. T$_{\mathrm{Day-Jupiter}_\mathrm{comb}}$, $T_{\mathrm{Day-Jupiter}_\mathrm{beat}}$, $T_{(\mathrm{Io-Jupiter}_\mathrm{beat}-\mathrm{Day})_\mathrm{beat}}$; see Appendix \ref{sec:Appendix_SNR_LS_versus_sample} for more details on the ratio between the S/N(LS) peaks of these different periodicities)}, which can be used to reinforce detections.

In our observations of Jupiter, data were often collected when Io-induced radio emissions were expected. While one might think this biases the results, this approach is consistent with our simulations (where the signal was always present) and parallels current efforts in exoplanet studies, where observations are programmed for a maximum likelihood of detecting signals, such as during quadrature phases with exoplanets {\citep{2023pre9.conf03091L} using prediction tools (e.g., ExPRES, \citet{2019A&A...627A..30L}; PALANTIR, \citet{2023pre9.conf03092M}; phase prediction, \citet{2025_ZHANG_Xiang_HD189733})}.

This technique will be applied to {a blind search} for exoplanetary or star--planet interaction signals in NenuFAR radio telescope data. {It will be useful to run it on detected radio signals \citep[several for a single source; e.g.][]{2023ApJ...953...65Z, 2025_AAASZhang, 2025A&A...695A..95Z, 2025NatAs...Tasse} to help determine possible origins by looking at the period organisation of the detected signals.}
{Depending on the configuration of the star--exoplanet system and the geometry of the observation, the detection of the signal \citep[depending on the orientation of the magnetic field with respect to the observer; e.g.][]{2023pre9.conf03091L} and the periodicities will be affected.
If a signal is detected at the orbital period of the exoplanet (or at the beat period between the star and the exoplanet), this would indicate an emission produced by the SPI; if the magnetic field of the star is dipolar or axisymmetric (i.e. aligned with the rotation axis), most of the LS power will be in the exoplanet's orbital period. On the other hand, if the magnetic field is either non-dipolar or non-axisymmetric (e.g. tilted, offset), most of the LS power will be in the beat period between the star and the exoplanet. More precisely, once the detection is done, the study of the individual bursts and in particular the variation of the maximal frequency of the emissions will inform us of the maximal magnetic-field value along the magnetic-field lines connected to the exoplanet, and therefore of the tilt, offset, and potential magnetic anomalies \citep{2011A&A...531A..29H}. A signal detected at the exoplanet's period of rotation would suggest an auroral emission from the exoplanet itself; a signal detected at the star's period of revolution would {indicate} auroral emission from the star itself; and, finally, no periodicity could point towards auroral emission from the star in the form of a hot spot \citep[sudden stellar eruptions; e.g.][]{2023ApJ...953...65Z, 2025_AAASZhang, 2025A&A...695A..95Z}.}

\begin{acknowledgements}
   C.L., A.L., P.Z., L.L. acknowledge funding from the ERC under the European Union's Horizon 2020 research and innovation program (grant agreement N$^\circ$ 101020459—Exoradio, \href{https://doi.org/10.3030/101020459}{doi: 10.3030/101020459}). The French authors acknowledge support from CNES and from CNRS/INSU programs of Planetology (PNP) and Heliophysics (ATST). The authors thank the anonymous referee for his meticulous work. Finally, C.L. would like to thank Q. Duchene and J. Morin for their help on getting information on ZDI and the detection limits, and E. Berriot for preliminary discussions on comparisons between periodicity detection techniques.
\end{acknowledgements}

\section*{Data availability statement}
Pre-processed data from the NenuFAR Key Project 07 observations of Jupiter used in this article can be accessed at \url{https://doi.org/10.25935/7a1s-rf17} \citep{2025NenuFAR_KP07_Jupiter_dataset_preprocessed_LS}. Unprocessed NenuFAR data are available upon request to PIs of key projects. The pipeline used to create this dataset can be accessed at \url{https://doi.org/10.5281/zenodo.15065695} \citep{louis_2025_15078360}.
The NDA dataset are available at \citet{2021NDA_data_collection} and the catalogue at \citet{vizier:J/A+A/604/A17}.

\bibliographystyle{aa}
\bibliography{bibliography}

\begin{appendix}
   \section{{Evolution of the Lomb--Scargle Periodogram as a Function of the Number of Samples}}
   \label{sec:Appendix_SNR_LS_versus_sample}

   {This appendix describes the evolution of the signal-to-noise Ratio (S/N) of the Lomb--Scargle (LS) Periodogram --denoted S/N(LS)-- as a function of the number of samples {$N$} included in the LS analysis. {For each LS periodogram, the S/N(LS) is calculated as S/N(LS$_\mathrm{N}$)~=~LS$_\mathrm{N}$/standard deviation (LS$_\mathrm{N}$)}. Figure \ref{fig:Figure_SNR_LS_over_samples_simulation} shows the evolution for the simulated signal (diluted into noise by a factor $K=8$). One can see in the top panel that it takes $N \simeq 20$ samples for the maximal peak in the LS periodogram to stabilize along the value of $T_\mathrm{sine}$. Note that for a dilution factor $K = 15$ (as in Figure \ref{fig:regularly_spaced_sinus_diluted}), it takes $N \simeq 55$ samples, and for a dilution factor $K=2$, it takes $N = 5$ samples. In the bottom panel one can see the evolution of the S/N(LS) of the associated peak as a function of the number of samples included in the LS analysis. It is observed that the S/N(LS) evolves as the square root of the number of samples (red curve). The black dashed lines highlight the sample number $N = 170$ for later comparison with the S/N(LS) evolution of the real data).}

   \begin{figure}[h]
      \centering
      \includegraphics[width=\columnwidth]{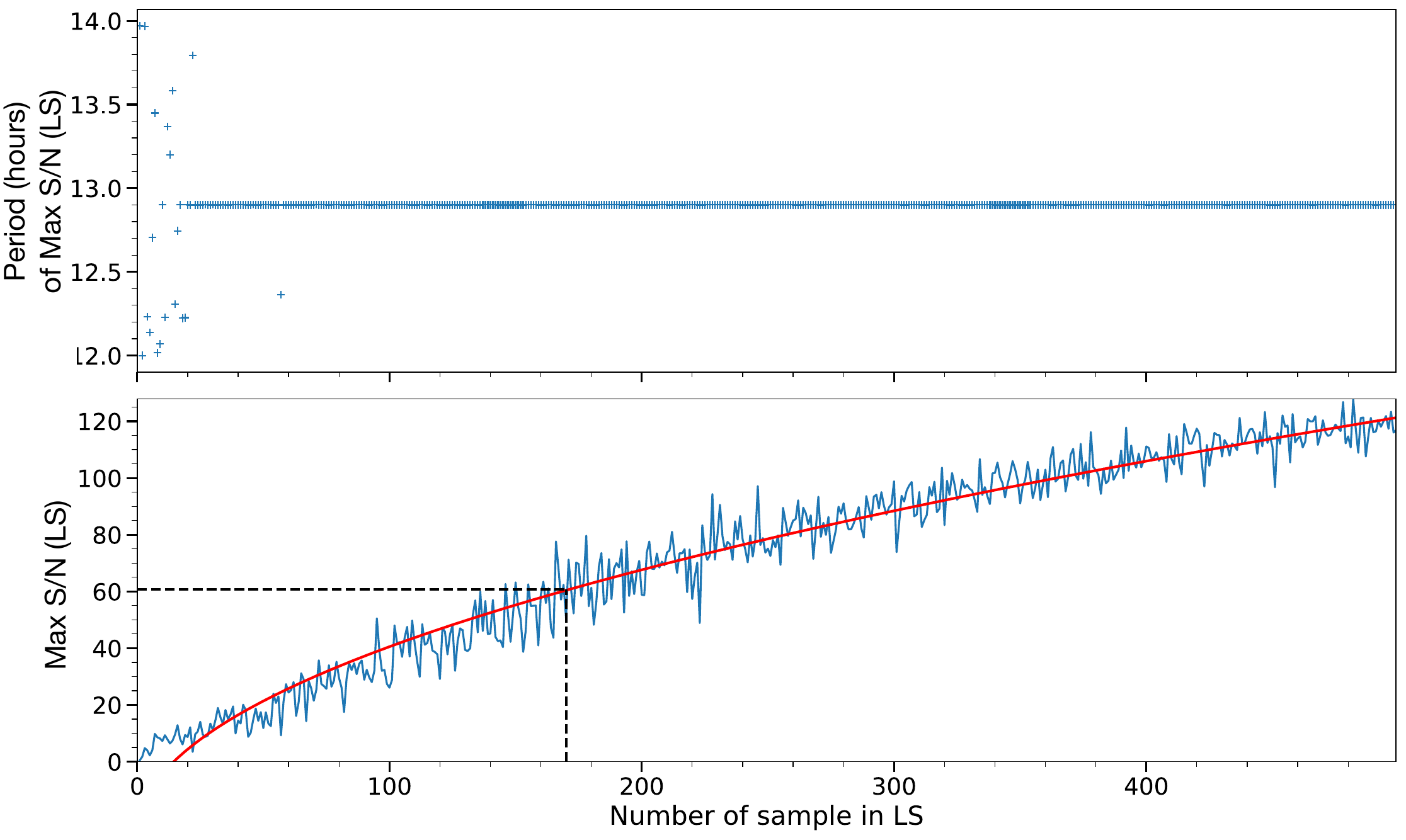}
      \caption{{Evolution as a function of the number of simulated observation samples inserted into the LS analysis of (top panel) the value of the maximum peak of the LS periodogram (in hours) and (bottom panel) of the S/N (of the LS periodogram) of the associated peak in the LS. The red curve is a simple fit $y= a \times \sqrt(x) + b$, with $x$ the number of samples, $a=5.188$ and $b = -14.545$. Here the simulated signal is diluted into noise, with a standard deviation $\sigma_\mathrm{normal~distribution} = K \times \sigma_\mathrm{sine}$ with $K=8$.. The black dashed lines highlight sample $N = 170$ corresponding to the number of observation samples in the real data (see Figure \ref{fig:Figure_SNR_LS_over_samples_data})}.}
      \label{fig:Figure_SNR_LS_over_samples_simulation}
   \end{figure}

   {Figure \ref{fig:Figure_SNR_LS_over_samples_data} displays the same as Figure \ref{fig:Figure_SNR_LS_over_samples_simulation} but for the real data. The different colors display different periodicities (see caption). {The ``Period of Max S/N(LS)'' is calculated for each periodicities at $\pm 2$~hours around the expected periodicity of interest (grey dashed--lines)}. {Depending on the periodicity studied, stability in the location of the Max S/N(LS) begins to be achieved from $ N \gtrsim 15$--$20$ (T$_\mathrm{Day-Jup-beat}$, T$_\mathrm{Jup}$, T$_\mathrm{Day}$), $N \gtrsim 45$ (T$_\mathrm{Io-Jup}$), $N \gtrsim 80$ (T$_\mathrm{Day-Jup-comb}$), or even $N \gtrsim 110$ (T$_\mathrm{Io}/2$)}. There is, of course, more variation than for the simulated signal, due to the beaming and the viewing effect of the radio emissions. Concerning the S/N(LS), one can see that, as for the simulated signal, it evolves as the square root of the number of samples.}

   {The bottom panel of Figure \ref{fig:Figure_SNR_LS_over_samples_data} allows for a more quantitative comparison of the S/N(LS) of the different periods. We will focus only on the last point of the distributions ($N=170$), corresponding to the values displayed in Figure \ref{fig:2D_periodogram}. The peak associated with $T_\mathrm{Jupiter}$ has an S/N(LS)~=~$47.4$;
   for $T_\mathrm{Jupiter}/2$: $9.0$;
   for $T_{\mathrm{Jupiter-Io}_\mathrm{beat}}$: $30.6$;
   for $T_\mathrm{Io}$: $3.8$ (not shown here);
   for $T_\mathrm{Io/2}: 18.5$;
   for $T_\mathrm{Day}$: $22.4$;
   for $T_\mathrm{Day/2}$: $15.3$ (not shown here);
   and for the beat and combination period:
    $T_{\mathrm{Jupiter-Io~beat~--~Io/2}_\mathrm{beat}}$: $4.4$ (not shown here);
    $T_{\mathrm{Jupiter-Day}_\mathrm{beat}}$: $25.2$;
    $T_{\mathrm{Jupiter-Day}_\mathrm{comb}}$: $27.2$;
     $T_{(\mathrm{Io-Jupiter}_\mathrm{beat}-\mathrm{Day})_\mathrm{beat}}$: $10.2$ (not shown here).}

   {The fact that the two highest S/N(LS) values are for $T_\mathrm{Jupiter}$ and $T_{\mathrm{Jupiter-Io}_\mathrm{beat}}$ highlights two things: (i) radio emissions are produced by Jupiter itself and by the interaction between Jupiter and its moon Io; (ii) the S/N(LS) of $T_{\mathrm{Jupiter-Io}_\mathrm{beat}}$ is higher than the S/N(LS) of $T_\mathrm{Io}$, proving that Jupiter's magnetic field is not axisymmetric. This is further highlighted by the S/N(LS) values higher for $T_\mathrm{Io/2}$ than for $T_\mathrm{Io}$.}
   
   {Due to the high S/N(LS) value for $T_\mathrm{Day}$, S/N(LS) of beat and/or combination periods between the three highest periods and $T_\mathrm{Day}$ are also detected.}

   \begin{figure}[h]
      \centering
      \includegraphics[width=\columnwidth]{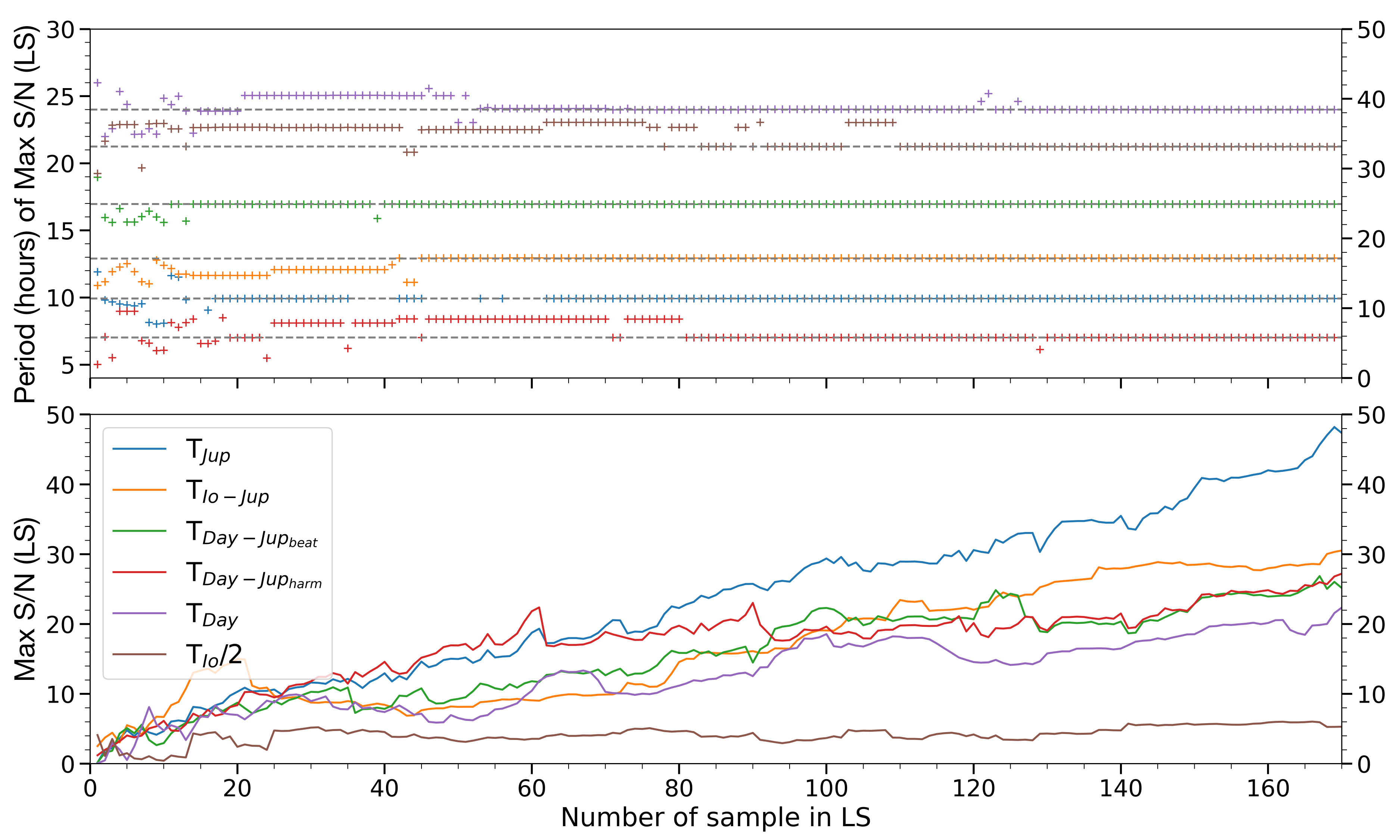}
      \caption{{Evolution as a function of the number of observation samples inserted into the LS analysis of (top panel) the values of the maximum peaks of the LS periodogram (in hours) --sought within $\pm2$~hours of the theoretical value (grey lines)-- and (bottom panel) of the S/N (of the LS periodogram) of the associated peak in the LS. The different colors correspond to the different periodicities detected in the LS periodogram: blue: $T_\mathrm{Jupiter}$; orange: $T_{\mathrm{Io-Jupiter}_\mathrm{ beat}}$; green: $T_{\mathrm{Day-Jupiter}_\mathrm{beat}}$; red: $T_{\mathrm{Day-Jupiter}_\mathrm{combination}}$; purple: $T_\mathrm{Day}$; brown: $T_\mathrm{Io/2}$.}}
      \label{fig:Figure_SNR_LS_over_samples_data}
   \end{figure}

   \section{Lomb--Scargle Analysis of the \citet{2017A&A...604A..17M} Nan\c cay Decameter Array Catalogue}
   \label{sec:Appendix_LS_NDA}

   \begin{figure}[h]
      \centering
      \includegraphics[width=\columnwidth]{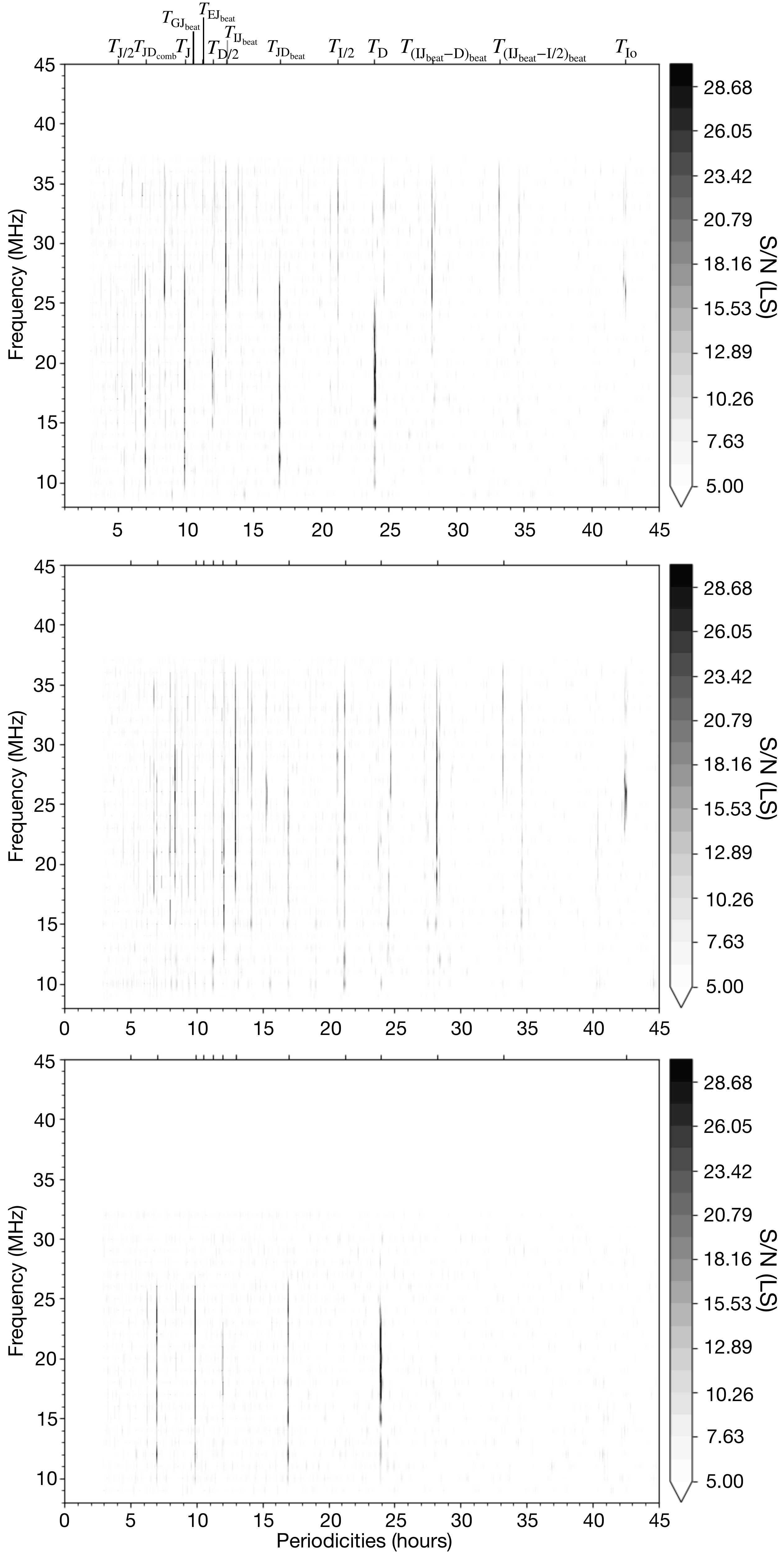}
      \caption{Signal to Noise ratio of the 2D LS Periodogram for (top panel) all emissions, (middle panel) Io--induced emissions, ({bottom} panel) non--Io emissions. {The} Y--axis {represents} the observed frequencies (from 8 to 45 MHz); {the} X--axis represents the LS periodicities. Periodicities of interest are indicated above the top y--axis.}
      \label{fig:Appendix_LS_NDA_SNR_2D}
   \end{figure}

   In this appendix are detailled the results of the LS analysis of the \citet{2017A&A...604A..17M} Nan\c cay Decameter Array Catalogue, ranging from  January 1990 to April 2020. In this catalogue, all observed Jovian emissions, were labeled with respect to their time--frequency morphology, their dominant circular polarization and maximum frequency. They are separated {into} Io and non--Io induced radio emissions. Figure \ref{fig:Appendix_LS_NDA_SNR_2D} {displays} the S/N of the 2D LS Periodograms for all emissions (top panel) and separated {into} Io--induced (middle panel) and non--Io emissions (bottom panel). For Io--induced emissions, high S/N {is} clearly visible at all periodicities related to Io ($T_\mathrm{Io}$, $T_\mathrm{Io/2}$, $T_\mathrm{{beat}~Io/2-IJ}$, $T_\mathrm{{beat}~D-IJ}$, $T_\mathrm{{beat}~IJ}$). For non--Io--induced emissions, high S/N {is} visible for periodicities linked to either Jupiter ($T_\mathrm{J}$) or the day ($T_\mathrm{D}$, $T_\mathrm{D/2}$), or {combinations} of both ($T_\mathrm{{beat}~DJ}$, $T_\mathrm{comb~DJ}$). These {periodicities} are detected at lower {observed} {frequencies} (up to $26$~MHz only) than Io--induced {emissions}, due to magnetic field topology. No clear peak {is} visible for Europa and Ganymede, which is also not surprising {as} they are not that often visible in the data, even if {they} were statistically detected using this catalogue \citep{2017pre8.conf...45Z,2018A&A...618A..84Z,2022A&A...665A..67J}, but after having {been} sorted by position of the moons as a function of the observer. 
   \end{appendix}
\end{document}